\documentclass{article}

\usepackage{iclr2027_conference,times}
\iclrpreprintcopy
\usepackage[T1]{fontenc}
\usepackage[utf8]{inputenc}
\usepackage{amsmath,amssymb}
\usepackage{booktabs}
\usepackage{siunitx}
\usepackage{graphicx}
\usepackage{float}
\usepackage{microtype}
\usepackage{tabularx}
\usepackage{array}
\usepackage{listings}
\usepackage[section]{placeins}
\usepackage{flafter}
\usepackage{afterpage}
\usepackage{url}
\usepackage{wrapfig}
\usepackage{algorithm}
\usepackage{algpseudocode}
\usepackage{fvextra}
\usepackage{multirow}
\usepackage[hyperfootnotes=false]{hyperref}

\usepackage{amsmath,amsfonts,bm}

\def\eqref#1{equation~\ref{#1}}

\def\1{\bm{1}}

\DeclareMathAlphabet{\mathsfit}{\encodingdefault}{\sfdefault}{m}{sl}
\SetMathAlphabet{\mathsfit}{bold}{\encodingdefault}{\sfdefault}{bx}{n}

\newcolumntype{Y}{>{\raggedright\arraybackslash}X}
\newcommand{\equalcontrib}{\textsuperscript{*}}
\newcommand{\projectlead}{\textsuperscript{\ensuremath{\dagger}}}
\newcommand{\correspondingauthor}{\textsuperscript{\ensuremath{\ddagger}}}

\title{\textsc{ReCast}: Contract-Preserving Protection for Fixed-Interface Multimodal Reasoning}
\author{%
  \parbox{0.94\textwidth}{\centering
    Bingchen Pei\textsuperscript{1}\equalcontrib,
    Lichong Chen\textsuperscript{1}\equalcontrib,
    Bingxi Zhao\textsuperscript{1}\equalcontrib\projectlead,
    Ziang Wu\textsuperscript{1}\equalcontrib,
    Sirui Wang\textsuperscript{1},\\
    Min Zhang\textsuperscript{2}, Yanhao Chen\textsuperscript{1},
    Qingxu Liu\textsuperscript{1},
    Qiang Gao\textsuperscript{1}, Chang-Tien Lu\textsuperscript{2},
    Bo Gao\textsuperscript{1}\correspondingauthor\\[0.75em]
    \small\textsuperscript{1}Beijing Jiaotong University\\
    \small\textsuperscript{2}Virginia Polytechnic Institute and State University\\[0.45em]
    {\mdseries\small\texttt{\{bogao, bingxizhao\}@bjtu.edu.cn}}
  }%
}
\date{}
\hypersetup{
  colorlinks=false,
  pdfborder={0 0 1},
  linkbordercolor={1 0 0},
  citebordercolor={0 1 0},
  urlbordercolor={0 1 1},
  pdftitle={ReCast: Contract-Preserving Protection for Fixed-Interface Multimodal Reasoning},
  pdfauthor={Bingchen Pei, Lichong Chen, Bingxi Zhao, Ziang Wu, Sirui Wang, Min Zhang, Yanhao Chen, Qingxu Liu, Qiang Gao, Chang-Tien Lu, Bo Gao}
}

\begin{document}
\maketitle
\begingroup
\renewcommand{\thefootnote}{}
\makeatletter
\long\def\@makefntext#1{\noindent#1}
\makeatother
\footnotetext{%
  \equalcontrib\,Contributed equally\\
  \projectlead\,Project lead\\
  \correspondingauthor\,Corresponding author
}
\endgroup
\setcounter{footnote}{0}

\begin{abstract}
Remote multimodal models offer strong numerical reasoning capabilities over charts and speech, but sending private inputs risks exposing sensitive content. Text-only sanitization cannot directly satisfy fixed media interfaces, while identity anonymization leaves the underlying task content exposed. We introduce \textbf{\textsc{ReCast}}, an agentic plug-in framework that replaces source-specific content while preserving task-relevant relations and the required input modality. \textbf{\textsc{ReCast}} locally converts inputs into a \textbf{shared textual evidence-query record}, jointly rewrites entities and topics with a distilled 4B model, and substitutes values through a locally invertible, \textbf{role-aware numerical map}. A \textbf{reconstruction agent} generates and validates the required media from the protected record. The remote solver returns a program whose protected operands are restored locally before execution. On 4,000 held-out ChartQA and NMSQA examples, \textbf{\textsc{ReCast}} achieves \textbf{75.10}\% accuracy, retaining \textbf{92.43}\% of unprotected remote accuracy, while a model-based audit flags source-content leakage in \textbf{7.95}\% of solver-bound requests. It outperforms all evaluated local baselines, preserving the benefit of remote reasoning while reducing source-content exposure under existing media interfaces.
\end{abstract}

\section{Introduction}

\begin{figure}[!t]
\centering
\includegraphics[width=\linewidth]{\detokenize{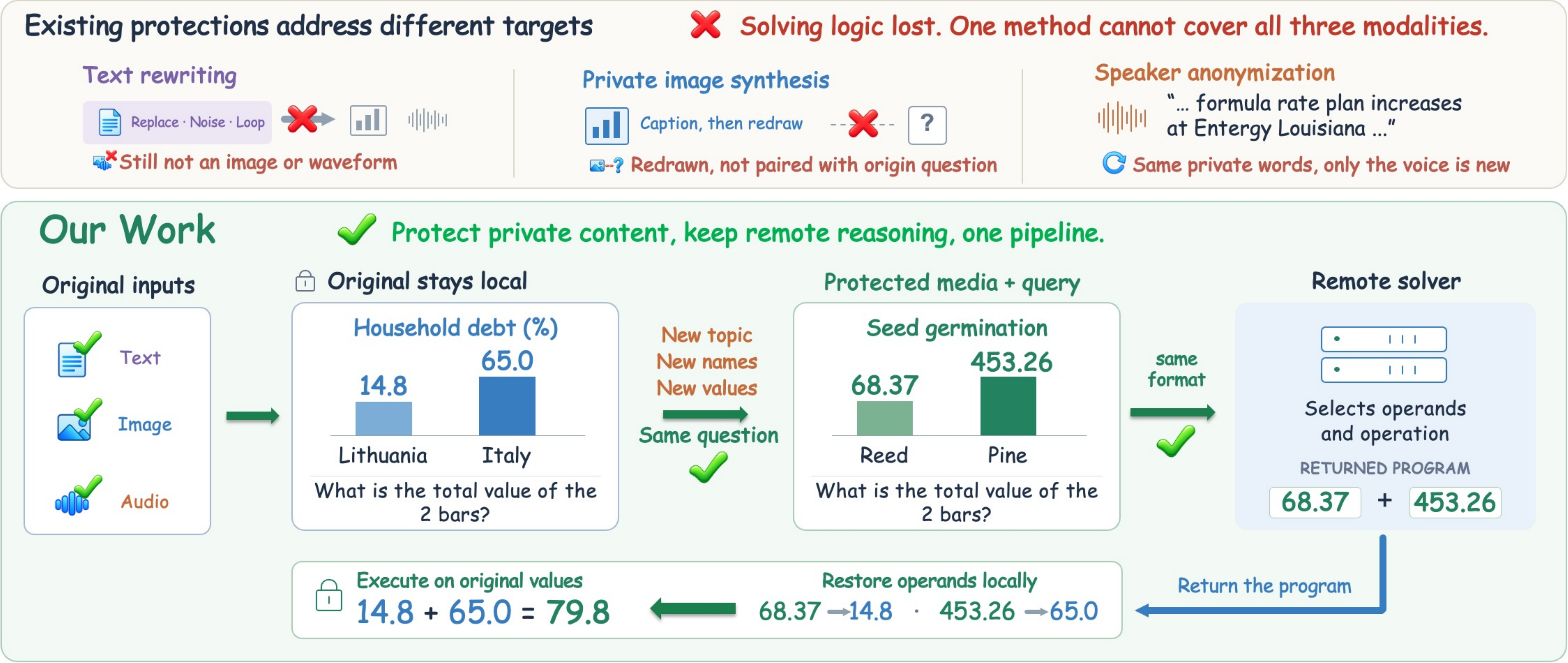}}
\caption{\textbf{\textsc{ReCast}: replacing source content while preserving task structure.}
\textbf{Top:} The illustrated alternatives address different targets:
text rewriting does not directly provide the required media,
dataset-level image synthesis is not paired with a specific query,
and speaker anonymization retains spoken content.
\textbf{Bottom:} \textbf{\textsc{ReCast}} jointly replaces topics, entities, and values
across evidence and query while retaining task-relevant relations
and the required modality.
The remote solver returns a program over protected operands;
the client restores the original operands locally before executing
the program to recover the answer.
The chart example illustrates a unified protection framework
for text, charts, and speech.}
\label{fig:teaser}
\end{figure}

Local-remote collaboration allows users to keep private data on their devices while leveraging a more capable remote model. This setting has been studied extensively for text. Before a request is released, a local model can rewrite it through differentially private rewriting \citep{meisenbacher2024dpmlm,thareja2025dpfusion}, adversarial anonymization \citep{duan2026rlaa}, or privacy-conscious delegation \citep{hartmann2024help,siyan2025papillon}. For numerical reasoning, \citet{zhang2026collaborative} shift the topic of a textual instance, substitute its numbers, and then recover the original answer locally from a program returned by the remote model.

However, much private evidence is not textual. Business reports present their figures in charts, while meetings and customer calls convey theirs in recordings. Reasoning over such evidence requires strong multimodal perception and numerical computation, areas in which on-device models still lag far behind remote ones. These remote models are typically accessed through service interfaces with fixed request schemas: a chart-analytics endpoint expects an image, and a spoken question-answering endpoint expects a waveform. Protection must therefore extend beyond text to the media these services consume.

Existing multimodal protections fall short of this need. Visual instruction rewriting \citep{mishra2025revision} and symbolic scene representations \citep{bara2022privatevqa} convert the input into text, which media interfaces do not accept. Media-level protections preserve the modality but target different threats: speaker anonymization conceals voice identity \citep{meyer2022anonymization}, adversarial perturbation prevents unauthorized speech recognition \citep{jin2025whispering}, and learned image transformations hide visual privacy attributes from multimodal models \citep{hui2026visual}. In each of these media-level methods, the content itself, such as which entity has which value in a chart or in speech, reaches the remote model intact. Numerical tasks impose a further requirement: evidence and query must be transformed consistently, and the transformed quantities must still support both the requested computation and the recovery of the original answer.

\textbf{\textsc{ReCast}} closes this gap by separating what a task needs from what the source reveals. The task needs relations, namely which quantity belongs to which entity, which pair the query compares, and which operation it requests, whereas the source contributes entities, topics, wording, and values. \textbf{\textsc{ReCast}} is an agentic plug-in pipeline built around a shared textual evidence-query record (Fig.~\ref{fig:teaser}). \emph{Evidence capture} plugs in a module for the input modality and locally converts the chart or recording into this record. A \emph{rewrite module} jointly replaces source-specific content across evidence and query while preserving their relations, and maps the numbers under a per-request map whose inverse remains local. A \emph{modality reconstruction agent} plugs in a generator for the interface contract and renders the protected record into the required modality through a generate-review-repair loop before anything reaches the solver. The remote solver returns a Python program over marked protected operands, which the local side restores and executes on the original values, extending the recovery scheme of \citet{zhang2026collaborative} to media inputs.

The main contributions of this work are as follows:

\begin{itemize}
\item We propose \textbf{\textsc{ReCast}}, an agentic plug-in pipeline that links evidence capture, rewriting, and modality reconstruction through a white-box, model-independent evidence-query record, so that per-modality and per-interface modules plug in without retraining.
\item We design a rewrite module in which a distilled local 4B Rewriter jointly changes names and domain while fixing relations, and a role-aware numerical map with local inversion enables exact answer recovery.
\item On 4,000 held-out ChartQA and NMSQA examples, \textbf{\textsc{ReCast}} keeps 92.43\% of unprotected accuracy with only 7.95\% solver-bound source-content leakage. Distillation adds 6.40 accuracy points and cuts leakage from 39.75\% to 7.95\%, and on ChartQA, reconstructed media slightly outperforms direct protected-text release by 0.40 points.
\end{itemize}

\section{Related Work}
\label{sec:related_work}

\subsection{Text privacy}
Text privacy methods range from word-level sanitization to document rewriting. SanText and CusText replace words with semantically related alternatives under differential privacy~\citep{yue2021santext,chen2023custext}, while DP-Prompt and DP-MLM use language models to produce coherent protected text~\citep{utpala2023dpprompt,meisenbacher2024dpmlm}. DP-Fusion instead limits the influence of sensitive tokens by combining generation distributions conditioned on different versions of the input~\citep{thareja2025dpfusion}. Beyond explicit identifiers, contextual clues can also reveal personal attributes~\citep{staab2024beyond}. Adversarial anonymization uses such inference attacks to guide text revisions, with RLAA adapting this approach to local models~\citep{staab2025anonymizers,duan2026rlaa}.

Recent methods also account for task utility and semantic structure. STAMP considers both task relevance and privacy sensitivity when allocating privacy budgets, while DP-ST uses semantic triples to guide private document generation~\citep{tian2026stamp,meisenbacher2025dpst}. These developments are relevant to our goal of preserving useful relations while replacing source content. Our setting adds two coupled requirements: evidence and query must remain aligned after numerical substitution, and the result must be usable through a fixed media interface. Text protection alone does not provide the required chart or waveform. \textsc{ReCast} therefore uses text as an editable intermediate shared across modalities, then reconstructs the protected record in the format the interface accepts.

\subsection{Local-remote collaboration}
Local-remote collaboration uses a local model to control what a stronger remote model sees. \citet{hartmann2024help} study how a model can seek help from another without revealing private information. CoGenesis and PAPILLON coordinate local and remote generation while limiting the private context exposed to the remote model~\citep{zhang2024cogenesis,siyan2025papillon}, whereas InferDPT combines local perturbation, remote generation, and local extraction~\citep{tong2025inferdpt}. More recent work makes delegation sensitive to the task and its privacy risks: Privacy-R1 learns which text to transmit, Veilmind conditions sanitization on user intent, and privacy-risk estimation accounts for contextual disclosures beyond direct identifiers~\citep{hui2026privacyr1,liu2026veilmind,siyan2026beyonddirect}.

The closest work is that of \citet{zhang2026collaborative}, who shift the topic of a textual numerical-reasoning instance, substitute its numbers, and recover the original answer locally from a program returned by the remote model. Separating program generation from execution is also central to program-aided reasoning~\citep{gao2023pal}. \textsc{ReCast} extends this recovery scheme to media inputs and outputs. A shared textual evidence-query record connects modality-specific evidence capture, joint rewriting, and reconstruction. The record makes cross-field references explicit, while reconstruction checks whether the released chart or speech still represents the protected evidence. The distinction is not simply the use of a multimodal solver: the released artifact must satisfy the media interface and retain the relations needed for local answer recovery.

\subsection{Multimodal and media privacy}
Privacy risks also arise from information embedded in media: vision-language models can infer personal attributes even from images without a direct depiction of the individual~\citep{tomekce2024attributes}. One response is to replace the raw input with an abstract representation. Symbolic scene representations support privacy-preserving visual question answering~\citep{bara2022privatevqa}, and ReVision rewrites image-dependent instructions into self-contained text~\citep{mishra2025revision}. Outside privacy protection, DePlot similarly separates chart perception from language reasoning through plot-to-table translation~\citep{liu2023deplot}. These approaches motivate an explicit intermediate representation, but their textual or symbolic outputs cannot directly replace an image or waveform required by a fixed media interface.

\begin{figure}[!t]
\centering
\includegraphics[width=\linewidth]{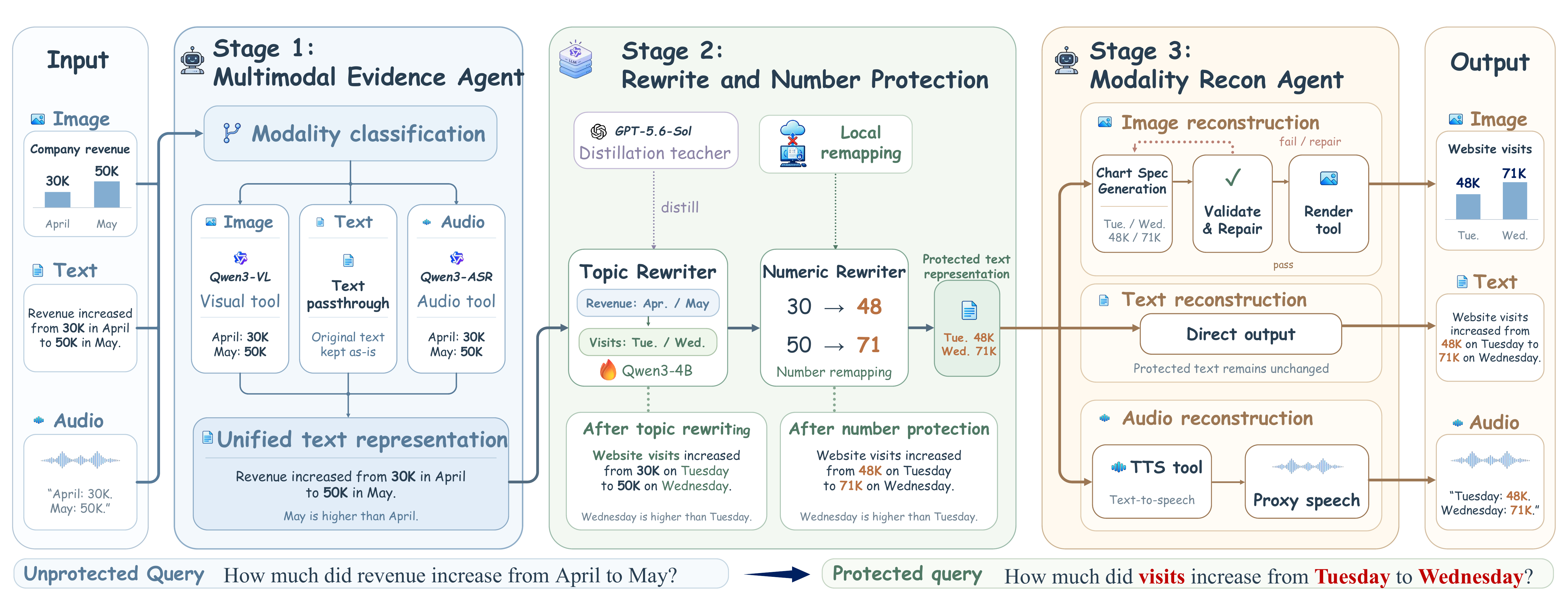}
\caption{\textbf{Overview of \textsc{ReCast}.} The pipeline first converts multimodal inputs into a unified textual representation (Sec.~\ref{sec:evidence-capture}), then applies topic rewriting and local numerical remapping to protect entities and numbers (Sec.~\ref{sec:rewrite-module}). The protected text is subsequently reconstructed into a surrogate in original modality, which regenerates charts through ChartSpec-based rendering and validation, passes through text directly, and synthesizes audio as proxy speech (Sec.~\ref{sec:modality-recon}). An example in the figure shows this pipeline.}
\label{fig:overall-pipeline}
\end{figure}

Media-level methods retain the modality but target different aspects of privacy. Speaker anonymization changes voice identity while retaining linguistic content~\citep{meyer2022anonymization}, SpeechShield masks sensitive speech segments before cloud recognition~\citep{benazir2025speechshield}, and learned image transformations protect visual privacy attributes from black-box multimodal models~\citep{hui2026visual}. SPTI is particularly related in using private textual intermediaries to synthesize differentially private images~\citep{wang2025spti}. It releases a synthetic dataset, however, rather than a protected media-query pair tied to a particular computation. \textsc{ReCast} instead replaces source-specific entities, topics, and values for each request while retaining the evidence-query relations needed for numerical reasoning. Protection comes from transforming the shared record; reconstruction provides interface compatibility, and the inverse numerical map remains local for answer recovery.

\section{Method}
\label{sec:method}

As shown in Fig.~\ref{fig:overall-pipeline}, \textsc{ReCast} separates source-specific content from task-relevant relations through a shared textual evidence-query record. It captures evidence locally (Sec.~\ref{sec:evidence-capture}), jointly rewrites its semantic and numerical content (Sec.~\ref{sec:rewrite-module}), and reconstructs the protected record in the required modality (Sec.~\ref{sec:modality-recon}). The remote solver returns a program whose operands are restored locally before execution.

\subsection{Task and Framework Overview}
\label{sec:overall-pipeline}

Given private evidence $X$ and a textual query $q$, let $m,r\in\{\mathrm{text},\mathrm{image},\mathrm{audio}\}$ denote the source and accepted release modalities, respectively. We primarily consider $r=m$, using another release modality only when the interface permits it. The goal is to reduce exposure of source-specific entities, topics, wording, and values while retaining the evidence-query associations needed for local answer recovery.

Let $E_m$ be the local evidence-capture operator, $D=E_m(X,q)$ its normalized textual record, $R_\theta$ the semantic rewriter with trainable adapter parameters $\theta$, and $H_h$ the numerical transformation induced by a request-specific injective map $h$. For requests passing release validation,

\begin{equation}
\widetilde D=H_h(R_\theta(D)),\qquad
(\widetilde X,\widetilde q)=G_r(\widetilde D).
\label{eq:protected-pipeline}
\end{equation}

Here, $\widetilde D$ is the protected record; $\widetilde X$ and $\widetilde q$ are the released evidence in modality $r$ and its protected textual query. $G_r$ denotes reconstruction and validation; its local validation inputs $(X,q,D)$ and inverse map $h^{-1}$ are implicit arguments. During deployment, these inputs and the inverse map remain local. Remote reconstruction services receive only protected content and derived artifacts; the solver receives $(\widetilde X,\widetilde q)$. The leakage audit covers solver-bound requests, not all reconstruction-service inputs.

\subsection{Multimodal Evidence Capture}
\label{sec:evidence-capture}

Evidence capture makes task-bearing content explicit without answering the query. For charts, local, question-aware Qwen3-VL-8B-Instruct~\citep{bai2025qwen3vl} extracts category-value associations, legend-series correspondences, units, ordering, and queried visual attributes. Qwen3-ASR-1.7B~\citep{shi2026qwen3asr} transcribes speech, while a deterministic parser normalizes text and preserves field boundaries and numerical spans.

All branches share five fields: \texttt{id}, \texttt{modality}, \texttt{context}, \texttt{question}, and \texttt{choices}, with empty choices for numeric-answer tasks. Numerical expressions are normalized before rewriting to match repeated quantities across evidence and query. This explicit, model-independent interface makes content inspectable and allows capture plug-ins to be replaced independently, provided their outputs remain schema-compatible.

\subsection{Joint Rewriting and Numerical Protection}
\label{sec:rewrite-module}

\begin{wrapfigure}{r}{0.54\linewidth}
    \centering
    \vspace{-8pt}
    \includegraphics[width=\linewidth]{\detokenize{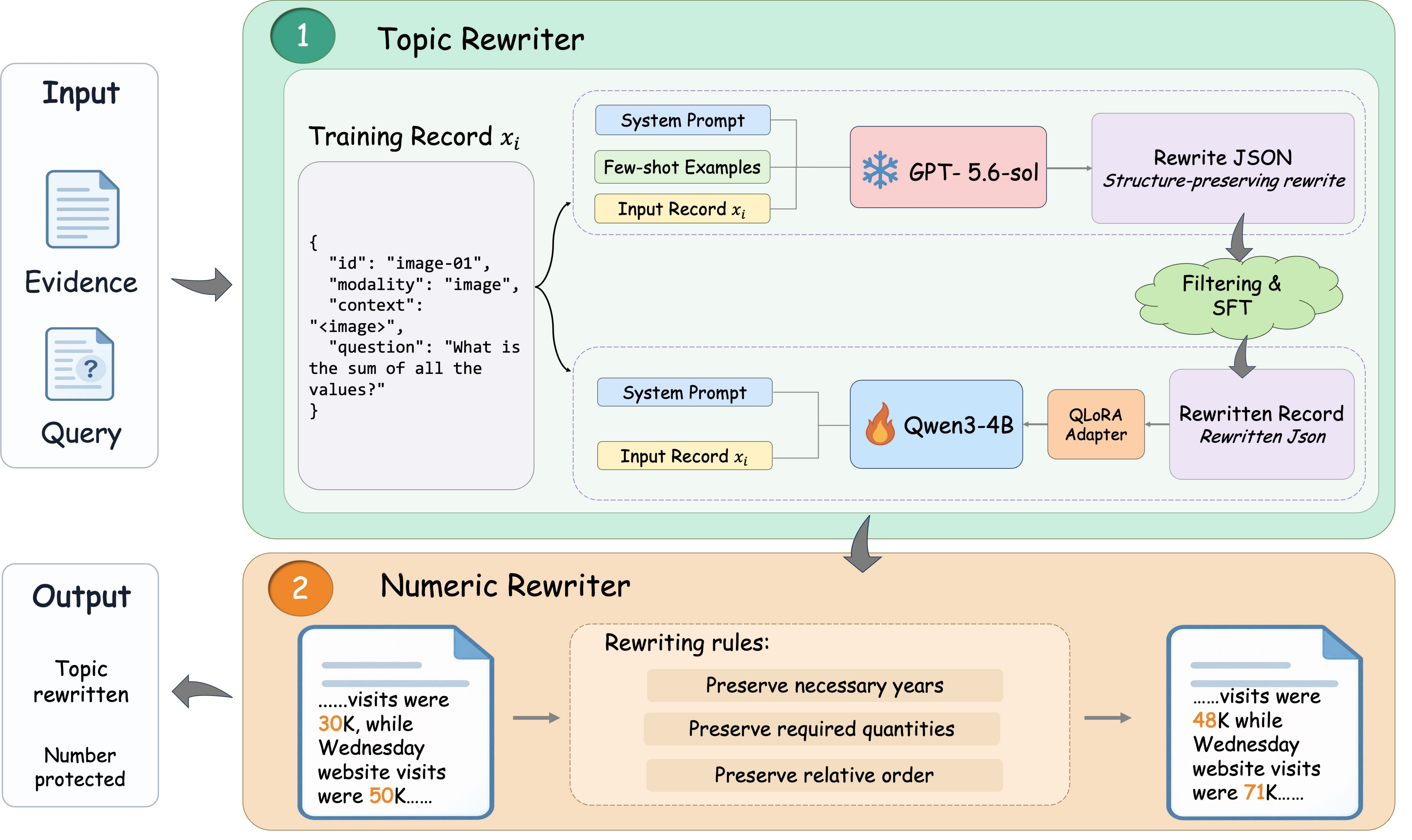}}
    \caption{\textbf{Joint rewriting and numerical protection.}
    A distilled local rewriter changes entities and topics consistently
    across evidence and query while retaining numerical literals.
    A separate request-specific map replaces numerical values under the
    supported relational constraints. Its inverse remains local for
    operand recovery.}
    \label{fig:rewrite-module}
    \vspace{-16pt}
\end{wrapfigure}

As shown in Fig.~\ref{fig:rewrite-module}, this stage protects the shared evidence-query record through two complementary transformations. A distilled local rewriter
jointly changes entities and topics while maintaining evidence-query alignment; a request-specific numerical map then substitutes values under explicit relational constraints. The inverse map remains local for restoring operands in the program returned by the remote solver.

\paragraph{Joint semantic rewriting.}

Evidence and query are rewritten together into a coherent, semantically distinct domain, retaining quantity assignments, cross-references, the requested operation, and answer type. Chart rewrites maintain category-value and legend-series correspondences, including queried visual attributes; speech rewrites maintain entity-quantity assignments, coreference, event order, and temporal or comparative links. Numerical literals remain unchanged, separating semantic rewriting from numerical substitution.

\paragraph{Distilling a local rewriter.}

We fine-tune Qwen3-4B~\citep{yang2025qwen3} with QLoRA~\citep{dettmers2023qlora} on offline GPT-5.6-Sol~\citep{openai2026gpt56} rewrites of training-side chart descriptions and speech transcripts. Schema and per-field numerical-literal filters yield 15,795 accepted pairs, split into training and development sets disjoint from evaluation data. Deployment uses the local student for rewriting. Let $\mathcal T$ be the training split and $Y$ the token sequence of the serialized teacher rewrite of $D$, before numerical substitution. We minimize

\begin{equation}
\mathcal L_{\mathrm{rewrite}}(\theta)
=-\sum_{(D,Y)\in\mathcal T}\sum_{t=1}^{|Y|}
\log p_\theta(Y_t\mid D,Y_{<t}).
\label{eq:rewrite-objective}
\end{equation}

Here, $Y_t$ is the $t$-th target token, $|Y|$ the sequence length, $Y_{<t}$ its preceding prefix, and $p_\theta$ the student's conditional next-token distribution. Loss applies only to output tokens. Inference outputs failing schema or numerical-literal checks are regenerated; semantic consistency is checked separately before release. Appendix~\ref{app:distillation} gives filtering and split details.

\paragraph{Numerical rewriting.}

Let $\mathcal V$ contain the distinct normalized values in the evidence and query. Each request uses an injective map $h:\mathcal V\rightarrow\widetilde{\mathcal V}$, where $\widetilde{\mathcal V}=h(\mathcal V)$, with local inverse $h^{-1}:\widetilde{\mathcal V}\rightarrow\mathcal V$. Let $\mathcal B_k\subseteq\mathcal V$ be the $k$-th magnitude band of remapped ordinary quantities and $\mathcal Y_g\subseteq\mathcal V$ the $g$-th contextual year group. Collision-free replacements are sampled and sorted within each band, while years in a group share an offset $\delta_g$:

\begin{align}
v<v' &\ \Longrightarrow\ h(v)<h(v'), \qquad v,v'\in\mathcal B_k,
\label{eq:within-band-order}\\
h(y)&=y+\delta_g, \qquad y\in\mathcal Y_g.
\label{eq:year-offset}
\end{align}

Here, $v,v'$ are ordinary values and $y$ is a year value. Repeated values receive the same replacement; non-year integers from 0 to 15 use identity assignments $h(v)=v$. The map preserves within-band order and within-group year intervals, but not general differences, ratios, or cross-class comparisons. Construction details appear in Appendix~\ref{app:numerical-map}.

\paragraph{Local operand recovery.}

Following~\citet{zhang2026collaborative}, the solver returns a Python program $\widetilde P$ with observed numerical operands marked as \texttt{protected(<number>)}. Let $\mathcal I_{h^{-1}}$ replace each marked operand $\texttt{protected}(h(v))$ with its original value $v$. The client computes

\begin{equation}
\widehat a=\operatorname{Exec}\!\left(
\mathcal I_{h^{-1}}[\widetilde P]\right),
\label{eq:local-recovery}
\end{equation}
where $\operatorname{Exec}(\cdot)$ denotes local program execution and $\widehat a$ is the predicted answer to $q$ on $X$. Recovery acts on operands, not on an answer computed from protected values. Correct recovery requires faithful evidence and correct operand and operation selection. Decisions based on unpreserved differences or ratios need not transfer unless deferred to program execution after restoration.

\subsection{Modality Reconstruction and Release Validation}

\begin{wrapfigure}{r}{0.54\linewidth}
    \centering
    \vspace{-6pt}
    \includegraphics[
        width=\linewidth
    ]{\detokenize{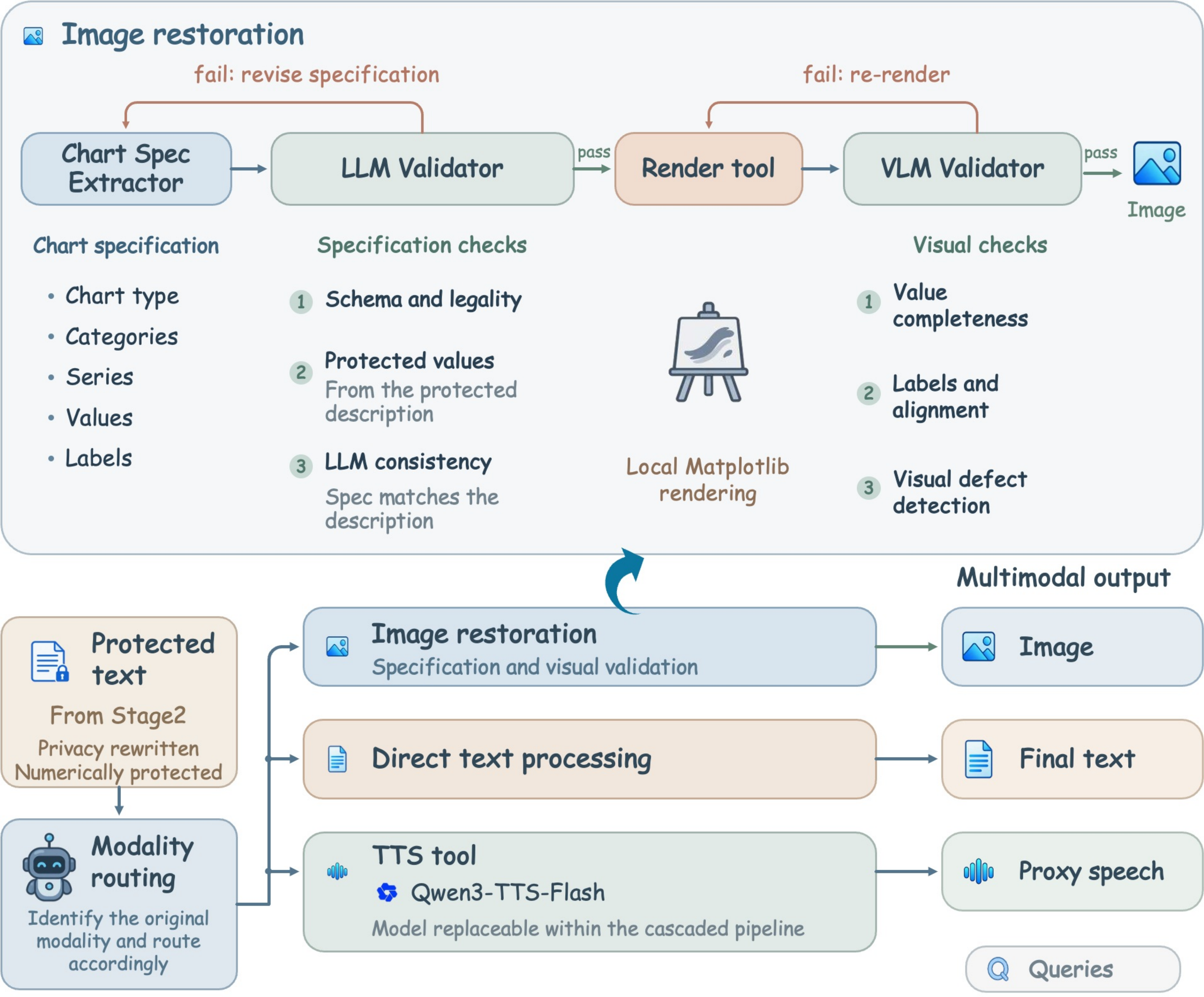}}
    \caption{\textbf{Modality reconstruction and release validation.}
    The interface selects the reconstruction plug-in. Charts use specification generation, local rendering, and review-guided repair; speech uses synthesis followed by transcription-based checks; text is serialized directly. Release requires both protected-record consistency and a local source-consistency check.}
    \label{fig:reconstruction-agent}
    \vspace{-34pt}
\end{wrapfigure}

\label{sec:modality-recon}

As shown in Fig.~\ref{fig:reconstruction-agent}, the reconstruction agent converts $\widetilde D$ into an interface-compatible artifact through a generate-review-repair loop. It aims to represent protected evidence faithfully, not to reproduce the original chart appearance or speaker identity.

\paragraph{Chart reconstruction.}

A remote code model translates protected evidence into a declarative ChartSpec containing the chart type, categories, series, values, labels, units, and colors. After local structure and value checks, fixed Matplotlib templates~\citep{hunter2007matplotlib} render the specification. A remote vision model reviews the image against the protected record, providing structured feedback for at most four repairs. Neither remote model receives the original chart. Failed repairs trigger a validated table fallback only when accepted by the interface; otherwise, the request is withheld.

\paragraph{Speech and text reconstruction.}
A remote text-to-speech (TTS) service synthesizes protected passages. The controller transcribes each candidate waveform, checks intelligibility, protected quantities, and entity references against the passage, and regenerates failures. For text-compatible interfaces, protected evidence is serialized directly with the protected query.

\paragraph{Two complementary release checks.}

Let $V_{\mathrm{protected}},V_{\mathrm{source}}\in\{0,1\}$ indicate whether a candidate passes the protected-record and local source-consistency checks, respectively, with $1$ for pass and $0$ for failure. The former checks rendered values, labels, units, associations, and queried attributes against $\widetilde D$ for charts, and transcribed quantities, references, and query-relevant content for speech. The latter aligns rewritten entities with source roles, restores mapped values, and compares assignments and dependencies with the original input. The release decision $V_{\mathrm{release}}\in\{0,1\}$ is

\begin{equation}
V_{\mathrm{release}}
=V_{\mathrm{protected}}\land V_{\mathrm{source}}.
\label{eq:release-gate}
\end{equation}

Here, $\land$ denotes logical conjunction; release occurs only when $V_{\mathrm{release}}=1$. These operational checks do not constitute a proof of semantic correctness or a formal privacy guarantee. Their coverage is detailed in Appendix~\ref{app:release-checks}.

\section{Experiments}
\label{sec:experiments}
\setlength{\textfloatsep}{6pt}
\setlength{\floatsep}{4pt}
\setlength{\intextsep}{5pt}
\setlength{\abovecaptionskip}{2pt}
\setlength{\belowcaptionskip}{0pt}
\renewcommand{\topfraction}{0.9}
\renewcommand{\textfraction}{0.05}
\setcounter{topnumber}{2}
This section evaluates \textsc{ReCast} along three main dimensions: overall reasoning utility and source-content protection, robustness beyond the primary evaluation setting, and the contribution of key pipeline components. We first describe the experimental setup (Sec.~\ref{sec:experiments-setup}), followed by the main results (Sec.~\ref{sec:experiments-main}), cross-dataset generalization (Sec.~\ref{sec:experiments-distill}), and stage-wise ablations (Sec.~\ref{sec:experiments-ablation}). Finally, we validate the LLM-based leakage judge against human annotations (Sec.~\ref{sec:experiments-judge}).

\subsection{Experimental Setup}
\label{sec:experiments-setup}

\paragraph{Datasets.}
We evaluate on four numeric-answer datasets. For in-distribution (ID)
evaluation, we use 2,000 held-out examples each from ChartQA
\citep{masry2022chartqa} and NMSQA \citep{lin2022dual}, disjoint from
the records used for Rewriter distillation. The distillation pool contains
9,370 ChartQA and 7,000 NMSQA records, yielding 15,795 accepted teacher
pairs split into 15,480 training and 315 validation examples.
For out-of-distribution (OOD) evaluation, we use 2,000 examples each
from ChartX \citep{xia2025chartxchartvlmversatile} and Spoken-MQA
\citep{wei2025spokenmathematicalreasoningbenchmarking}.
All evaluation IDs are fixed before execution
(Appendix~\ref{app:eval-data}). Unless otherwise stated, results are
reported on the ID sets; the OOD sets are used for the cross-dataset
evaluation in Sec.~\ref{sec:experiments-distill}.

\paragraph{Baselines.}
Because \textsc{ReCast} is a multi-stage pipeline, we compare it both end-to-end and at the stage at which each alternative operates.
\emph{End-to-end} \emph{Remote-native} sends the original media and query
to the same remote solver and provides the unprotected accuracy reference.
Gemma-4-E4B \citep{gemmateam2026gemma4} and Qwen2.5-Omni-3B
\citep{xu2025qwen25omni} answer the original inputs locally and serve as
zero-leakage references. On ChartQA, we additionally evaluate the Stage~1
capture model Qwen3-VL-8B as a standalone local solver.
\emph{At Stage~2,} we replace the Rewriter while keeping all other pipeline
components fixed. The untuned Qwen3-4B isolates the effect of distillation, while GPT-5.6-Sol,
the offline teacher that produces the distillation pairs, provides a
teacher-level reference. We also apply two text anonymizers to the same textual records: RLAA \citep{duan2026rlaa},
which we port to a Qwen3-4B backbone, and Eternis-Anonymizer-4B
\citep{eternis2025anonymizer}; both are evaluated as stage-level
references.
\emph{At Stage~3,} we consider two groups of alternatives. The first group
protects the media directly: PSIC \citep{shen2025psic} produces protected
images, SPTI \citep{wang2025spti} synthesizes differentially private images
from private textual intermediaries, and AudioShield
\citep{jin2025whispering} protects audio against speech recognition.
None of these methods restores the original numerical operands after
protected inference. They therefore do not yield answerable end-to-end QA
instances under our protocol, so we report their native exposure diagnostics
instead of QA accuracy. The second group isolates the reconstruction strategy: we replace the \textsc{ReCast} reconstruction
agent with two alternative agent harnesses, Codex CLI
\citep{openai2026codexcli} and DeepSeek Harness (DSH)
\citep{deepseekai2026harness}, which reconstruct the same protected
records. Stage~1 has no direct external counterpart and is examined
through a controlled ablation in Sec.~\ref{sec:experiments-ablation}.

\begin{table}[!t]
\centering
\setlength{\tabcolsep}{4pt}
\renewcommand{\arraystretch}{1.08}
\caption{
Stage~2 Rewriter comparison (a) and native exposure diagnostics of
protection references (b). Values are percentages;
$\Delta$: \textsc{ReCast} minus the reference (pp);
``--'': unavailable; $^{\ast}$: single-dataset value.
}
\label{tab:1}

\phantomsection\label{tab:1a}
\resizebox{\linewidth}{!}{%
\begin{tabular}{@{}lrrrrrr@{}}
\toprule
\multicolumn{7}{@{}l}{
\textbf{(a) Controlled Rewriter comparison}} \\
\addlinespace[3pt]

\multirow{2}{*}{Rewriter}
& \multicolumn{3}{c}{Accuracy $\uparrow$}
& \multicolumn{3}{c}{Source-content leakage $\downarrow$} \\
\cmidrule(lr){2-4}
\cmidrule(lr){5-7}
& ChartQA & NMSQA & Overall
& ChartQA & NMSQA & Overall \\
\midrule

Qwen3-4B (untuned, local)
& 66.40 & 71.00 & 68.70
& 37.90 & 41.60 & 39.75 \\

\textbf{\textsc{ReCast}} (distilled local 4B)
& 70.20 & 80.00 & 75.10
& 8.00 & 7.90 & 7.95 \\

GPT-5.6-Sol (offline teacher)
& 70.30 & 81.00 & 75.65
& 5.20 & 1.20 & 3.20 \\

\midrule

\quad$\Delta$ vs.\ untuned (pp)
& $+3.80$ & $+9.00$ & $+6.40$
& $-29.90$ & $-33.70$ & $-31.80$ \\

\quad$\Delta$ vs.\ teacher (pp)
& $-0.10$ & $-1.00$ & $-0.55$
& $+2.80$ & $+6.70$ & $+4.75$ \\

\bottomrule
\end{tabular}%
}

\medskip

\phantomsection\label{tab:1b}
\resizebox{\linewidth}{!}{%
\begin{tabular}{@{}lp{4.0cm}rrr@{}}
\toprule
\multicolumn{5}{@{}l}{
\textbf{(b) Protection references and task applicability}} \\
\addlinespace[3pt]

\multirow{2}{*}{Method}
& \multirow{2}{*}{Role / task applicability}
& \multicolumn{3}{c}{Reported exposure diagnostic} \\
\cmidrule(lr){3-5}
& & ChartQA & NMSQA & Overall \\
\midrule

\multicolumn{5}{@{}l}{\emph{Text-rewriting references}} \\

RLAA-Qwen3-4B (port)
& Stage-level reference
& 86.80 & 73.70 & 80.25 \\

Eternis-Anonymizer-4B
& Stage-level reference
& 96.70 & 93.20 & 94.95 \\

\midrule

\multicolumn{5}{@{}l}{\emph{Media-protection references}} \\

PSIC
& No end-to-end recovery
& 28.20 & -- & $28.20^{\ast}$ \\

SPTI
& No end-to-end recovery
& 0.00 & -- & $0.00^{\ast}$ \\

AudioShield
& No end-to-end recovery
& -- & 0.60 & $0.60^{\ast}$ \\

\bottomrule
\end{tabular}%
}
\end{table}
\begin{table}[!t]
\centering

\begin{minipage}[t]{0.48\linewidth}
\vspace{0pt}
\centering

\caption{
Accuracy (\%) of Stage~3 reconstruction implementations.
}
\label{tab:2}

\vspace{2pt}
\renewcommand{\arraystretch}{1.45}

\resizebox{\linewidth}{!}{%
\begin{tabular}{@{}lrrr@{}}
\toprule
Reconstruction & ChartQA & NMSQA & Overall \\
\midrule
Codex CLI harness & 52.00 & 78.00 & 65.00 \\
DeepSeek Harness (DSH) & 60.90 & 80.50 & 70.70 \\
\textbf{\textsc{ReCast} agent (Ours)} & \textbf{70.20} & \textbf{80.00} & \textbf{75.10} \\
\bottomrule
\end{tabular}%
}

\end{minipage}
\hfill
\begin{minipage}[t]{0.48\linewidth}
\vspace{0pt}
\centering

\caption{
ID and OOD results (\%) across image and audio modalities.
}
\label{tab:3}

\vspace{2pt}

\begin{tabular}{@{}lcccc@{}}
\toprule
& \multicolumn{2}{c}{Image} & \multicolumn{2}{c}{Audio} \\
\cmidrule(lr){2-3}
\cmidrule(lr){4-5}
Metric & ID & OOD & ID & OOD \\
\midrule
Acc. $\uparrow$
& 70.20 & 62.40
& 80.00 & 79.40 \\

Leakage $\downarrow$
& 8.00 & 1.40
& 7.90 & 5.90 \\
\bottomrule
\end{tabular}%

\end{minipage}

\end{table}

\paragraph{Experiments.}
We instantiate the three stages of \textsc{ReCast} as follows. In Stage~1,
Qwen3-VL-8B-Instruct \citep{bai2025qwen3vl} and Qwen3-ASR-1.7B
\citep{shi2026qwen3asr} convert chart and speech inputs, respectively, into
the shared textual record. In Stage~2, the local Rewriter is Qwen3-4B
\citep{yang2025qwen3} fine-tuned with QLoRA \citep{dettmers2023qlora} on
the accepted teacher pairs. In Stage~3, DeepSeek-V4 models
\citep{deepseekai2026deepseekv4} generate and review ChartSpecs, and
Qwen3-TTS-Flash \citep{hu2026qwen3tts} synthesizes speech.
Qwen3.5-Omni-Plus \citep{qwenteam2026qwen35omni} serves as the remote
solver in all remote settings, including Remote-native, so accuracy
differences are not attributable to the choice of solver. Following the recovery scheme of \citet{zhang2026collaborative}, we recover
the original answer locally by restoring the protected operands. All local
components run sequentially on a single NVIDIA RTX 4090; full
configurations are given in Appendices~\ref{app:components}
and~\ref{app:hyperparameters}.
We report task accuracy (Acc.), normalized accuracy (NAcc.), and
source-content leakage (Leak.). NAcc. is the ratio of a method's accuracy to
Remote-native accuracy on the same examples. Leak. is the fraction of
solver-bound requests flagged by a fixed LLM judge (validated against human
labels in Sec.~\ref{sec:experiments-judge}) for semantic, lexical,
entity-value correspondence, or source-linkability information; generic
reasoning structure is not counted as leakage. For tables that report an
Overall value, we use the arithmetic mean of the ChartQA and NMSQA values. No
example is filtered post hoc by accuracy or leakage, and answer
normalization and matching are specified in
Appendix~\ref{app:answer-matching}.

\begin{table}[!t]
\centering

\begin{minipage}[t]{0.48\linewidth}
\vspace{0pt}
\centering

\caption{
Stage~1 input-path ablation (accuracy, \%).
}
\label{tab:4}

\vspace{2pt}

\resizebox{\linewidth}{!}{%
\begin{tabular}{@{}lrrr@{}}
\toprule
Input path & Image acc. & Audio acc. & Overall acc. \\
\midrule
Joint input & 46.00 & 43.20 & 44.60 \\
Decoupled input & \textbf{65.90} & \textbf{65.80} & \textbf{65.85} \\
\bottomrule
\end{tabular}%
}

\end{minipage}
\hfill
\begin{minipage}[t]{0.48\linewidth}
\vspace{0pt}
\centering

\caption{
Stage~3 release-format ablation (accuracy, \%).
}
\label{tab:5}

\vspace{2pt}

\resizebox{\linewidth}{!}{%
\begin{tabular}{@{}lrrr@{}}
\toprule
Release format & ChartQA & NMSQA & Overall \\
\midrule
\textbf{\textsc{ReCast}}, protected media & 70.20 & 80.00 & 75.10 \\
\textbf{\textsc{ReCast}}, protected text & 69.80 & 83.90 & 76.85 \\
\bottomrule
\end{tabular}%
}

\end{minipage}

\end{table}

\subsection{Main Results}
\label{sec:experiments-main}

Our experiments compare three settings: local models operating on the original inputs, \textsc{ReCast} operating on protected media, and remote-native inference, in which the same remote solver receives the original media and query. The local zero-leakage references and the fully exposed remote-native reference are reported in Appendix Tab.~\ref{tab:main-results}. \textsc{ReCast} achieves 70.20\% accuracy (NAcc.~93.48\%) on ChartQA, 80.00\% (NAcc.~91.53\%) on NMSQA, and 75.10\% overall (NAcc.~92.43\%), with an overall solver-bound source-content leakage of 7.95\%. These results show that the protected pipeline retains most of the
reasoning utility of the unprotected remote solver while reducing
source-content exposure at the solver boundary.

Despite operating on protected inputs, \textsc{ReCast} also outperforms the local reference models. On ChartQA, it reaches 70.20\%, exceeding the standalone Qwen3-VL-8B capture model (62.80\%) by 7.40 points. On NMSQA, it reaches 80.00\%, exceeding Qwen2.5-Omni-3B (46.30\%) by 33.70 points.

Tab.~\hyperref[tab:1a]{\ref*{tab:1}(a)} next examines the effect of the Stage~2 Rewriter. Distillation brings \textsc{ReCast} close to its GPT-5.6-Sol teacher while substantially outperforming the untuned Qwen3-4B: it trails the teacher (75.65\%) by only 0.55 points, improves over the untuned model by 6.40 points, and reduces leakage from 39.75\% to 7.95\%. The two text anonymizers applied to the same textual records, reported in Tab.~\hyperref[tab:1b]{\ref*{tab:1}(b)}, leave source content in 80.25\% (RLAA) and 94.95\% (Eternis) of their outputs, an order of magnitude above the 7.95\% of \textsc{ReCast}. General-purpose anonymization thus leaves most source content in place, whereas the distilled Rewriter jointly replaces the entities and topics of the record. Among the methods reported in Tab.~\ref{tab:1}, \textsc{ReCast} is the only configuration for which we report end-to-end QA accuracy on both modalities.

For Stage~3, Tab.~\hyperref[tab:1b]{\ref*{tab:1}(b)} reports the native exposure diagnostics for PSIC, SPTI, and AudioShield described in Sec.~\ref{sec:experiments-setup}, rather than end-to-end QA accuracy. These diagnostics use method-specific protocols and are therefore not directly comparable with \textsc{ReCast}'s solver-bound leakage rate. The 0.00 reported for SPTI indicates that no source-content leakage was detected under its native exposure diagnostic. Appendices~\ref{app:psic}-\ref{app:audioshield} provide additional diagnostics, while Appendix~\ref{app:visualization-results} compares their outputs with those of \textsc{ReCast} and the two text anonymizers: Figures~\ref{fig:visualization-images} and~\ref{fig:visualization-audio} show selected image and audio cases, respectively, and the selected \textsc{ReCast} examples retain correct final answers.


Tab.~\ref{tab:2} isolates the Stage~3 reconstruction implementation by
passing the same protected examples through each reconstruction path. \textsc{ReCast} outperforms the Codex CLI harness~\citep{openai2026codexcli} by 18.20 points on ChartQA and 10.10 points overall, and outperforms DeepSeek Harness~\citep{deepseekai2026harness} by 9.30 points on ChartQA and 4.40 points overall. On NMSQA, DeepSeek Harness is 0.50 points higher, indicating that the advantage of \textsc{ReCast} comes mainly from chart reconstruction rather than holding uniformly across modalities.

\subsection{Cross-Dataset Evaluation}
\label{sec:experiments-distill}

Tab.~\ref{tab:3} evaluates \textsc{ReCast} with the distilled Qwen3-4B rewriter
under both in-distribution (ID) and out-of-distribution (OOD) settings to
test whether it maintains task accuracy and low source-content leakage on
unseen datasets. The ID evaluation uses ChartQA and NMSQA, where the full pipeline achieves 70.20\% accuracy on images and 80.00\% on audio. On the OOD datasets, ChartX and Spoken-MQA, accuracy reaches 62.40\% and 79.40\%, corresponding to drops of 7.80 and 0.60 points from the respective ID results. The substantially smaller OOD degradation on audio indicates stronger cross-dataset transfer
for spoken inputs, while image performance is more sensitive to shifts in
chart distribution. The measured leakage rates are 8.00\% and 1.40\% for the ID and OOD
image datasets, and 7.90\% and 5.90\% for the ID and OOD audio
datasets, respectively. Because the datasets differ, these values are
reported descriptively rather than interpreted as an effect of
distillation.





\subsection{Ablations}
\label{sec:experiments-ablation}

To test whether a shared textual representation is an effective intermediate
interface for multimodal evidence, we compare direct joint multimodal input
with the proposed decoupled pathway. Both paths use the same multimodal
model: the joint path rewrites the original input directly, whereas the
decoupled path first converts modality-specific evidence into the shared
textual representation and rewrites that representation. This is a Stage~1
path ablation under a shared model configuration, rather than a
reproduction of the full \textsc{ReCast} setup. Tab.~\ref{tab:4} shows that
decoupling raises overall accuracy from 44.60\% to 65.85\% (+21.25 points),
with gains on both image (+19.90) and audio (+22.60), supporting the
decoupled Stage~1 design.

To assess the effect of Stage~3 reconstruction, we keep the protected record fixed and compare releasing it as reconstructed media or directly as protected text. Tab.~\ref{tab:5} shows that protected text is 1.75 points higher overall (76.85\% versus 75.10\%), mainly due to a 3.90-point gain on NMSQA. On ChartQA, however, reconstructed images slightly outperform the protected
textual representation by 0.40 points. This indicates that reconstructing the
protected record into the required image modality can preserve downstream
utility in this setting, while also satisfying image-based service interfaces. For audio, reconstructed speech performs 3.90 points below protected text, indicating a modest utility loss introduced by speech synthesis. Nevertheless, for services that require waveform input, audio reconstruction remains necessary for interface-compatible end-to-end inference. Overall, the comparison isolates the effect of modality reconstruction from the task content already encoded in the shared protected record.


\subsection{Validity of the LLM Judge}
\label{sec:experiments-judge}

Source-content leakage cannot be measured by string matching alone: a
solver-bound request can reveal the source through paraphrased semantics,
entity-value correspondences, or cues that link it back to the original
input, whereas retaining generic reasoning structure does not constitute
leakage. We therefore rely on a fixed LLM judge for all leakage rates of
\textsc{ReCast} and its variants reported in
Secs.~\ref{sec:experiments-main}-\ref{sec:experiments-distill}. To assess whether the judge provides a reliable proxy for human assessment,
we compare its decisions with human labels on 400 annotated pairs. The judge agrees
with the human majority on 91.50\% of the pairs ($\kappa=0.83$), indicating agreement well beyond chance. The annotation protocol and calibration
details are given in Appendix~\ref{app:human-calibration}.

\section{Conclusion}

We presented \textsc{ReCast}, a contract-preserving protection framework
for numerical reasoning through fixed multimodal interfaces.
Its central insight is to replace source-specific content while
preserving the task relations and media format required by the service.
A shared evidence-query record decouples content protection from
modality, while local operand recovery separates remote reasoning
over protected evidence from computation on original values.
On ChartQA and NMSQA, \textsc{ReCast} retains 92.43\% of unprotected
remote accuracy, with a model-based audit flagging source-content
leakage in 7.95\% of solver-bound contexts.
With a distilled local 4B rewriter, end-to-end accuracy falls
just 0.55 percentage points below the offline teacher configuration.
The reconstruction agent improves chart accuracy over the evaluated
alternatives, while results on ChartX and Spoken-MQA support
transfer beyond the distillation datasets.
Together, these findings show that solver-bound source-content exposure
can be reduced while retaining much of the utility of remote numerical
reasoning and compatibility with existing media interfaces.

\section*{AI Use Statement}

Generative AI tools were used to assist with language editing, improving the clarity and organization of the manuscript, and providing feedback on presentation and experimental descriptions. They were also used to assist with code editing. All research ideas, methodological decisions, experimental results, and scientific claims were reviewed and verified by the authors. AI-generated suggestions were not treated as authoritative sources, and the authors take full responsibility for the final content of this work.

\section*{Reproducibility Statement}

Appendix~\ref{app:implementation} lists data construction, components, the mapping policy, and distillation data; Appendix~\ref{app:hyperparameters} lists training and decoding settings; Appendix~\ref{app:prompts} gives the prompt specifications. We will release the code, prompts, the distilled adapter, the accepted teacher traces, per-run manifests (selected sample IDs, offsets, and code, environment, and data hashes), and per-example request and audit logs.

\section*{Ethics Statement}

ChartQA and NMSQA are public benchmarks. The three annotators in the human calibration (Appendix~\ref{app:human-calibration}) are authors of this paper. The leakage judge receives original inputs only inside the evaluation workflow; in deployment, original inputs and inverse maps stay local. \textsc{ReCast} lowers the source content that reaches remote services, and sensitive deployments can combine it with contractual and access controls on those services.

\bibliographystyle{iclr2027_conference}
\bibliography{ref}

\begin{thebibliography}{48}
\providecommand{\natexlab}[1]{#1}
\providecommand{\url}[1]{\texttt{#1}}
\expandafter\ifx\csname urlstyle\endcsname\relax
  \providecommand{\doi}[1]{doi: #1}\else
  \providecommand{\doi}{doi: \begingroup \urlstyle{rm}\Url}\fi

\bibitem[Bai et~al.(2025)Bai, Cai, Chen, Chen, Chen, Cheng, Deng, Ding, Gao, Ge, Ge, Guo, Huang, Huang, Huang, Hui, Jiang, Li, Li, Li, Li, Lin, Lin, Liu, Liu, Liu, Liu, Liu, Liu, Lu, Luo, Lv, Men, Meng, Ren, Ren, Song, Sun, Tang, Tu, Wan, Wang, Wang, Wang, Wang, Xie, Xu, Xu, Xu, Yang, Yang, Yang, Yang, Yu, Zhang, Zhang, Zhang, Zheng, Zhong, Zhou, Zhou, Zhou, Zhu, and Zhu]{bai2025qwen3vl}
Shuai Bai, Yuxuan Cai, Ruizhe Chen, Keqin Chen, Xionghui Chen, Zesen Cheng, Lianghao Deng, Wei Ding, Chang Gao, Chunjiang Ge, Wenbin Ge, Zhifang Guo, Qidong Huang, Jie Huang, Fei Huang, Binyuan Hui, Shutong Jiang, Zhaohai Li, Mingsheng Li, Mei Li, Kaixin Li, Zicheng Lin, Junyang Lin, Xuejing Liu, Jiawei Liu, Chenglong Liu, Yang Liu, Dayiheng Liu, Shixuan Liu, Dunjie Lu, Ruilin Luo, Chenxu Lv, Rui Men, Lingchen Meng, Xuancheng Ren, Xingzhang Ren, Sibo Song, Yuchong Sun, Jun Tang, Jianhong Tu, Jianqiang Wan, Peng Wang, Pengfei Wang, Qiuyue Wang, Yuxuan Wang, Tianbao Xie, Yiheng Xu, Haiyang Xu, Jin Xu, Zhibo Yang, Mingkun Yang, Jianxin Yang, An~Yang, Bowen Yu, Fei Zhang, Hang Zhang, Xi~Zhang, Bo~Zheng, Humen Zhong, Jingren Zhou, Fan Zhou, Jing Zhou, Yuanzhi Zhu, and Ke~Zhu.
\newblock {Qwen3-VL Technical Report}, 2025.
\newblock URL \url{https://arxiv.org/abs/2511.21631}.

\bibitem[Bara et~al.(2022)Bara, Ping, Mathur, Thattai, {Rohith MV}, and Sukhatme]{bara2022privatevqa}
Cristian-Paul Bara, Qing Ping, Abhinav Mathur, Govind Thattai, {Rohith MV}, and Gaurav~S. Sukhatme.
\newblock {Privacy Preserving Visual Question Answering}, 2022.
\newblock URL \url{https://arxiv.org/abs/2202.07712}.

\bibitem[Benazir \& Lin(2025)Benazir and Lin]{benazir2025speechshield}
Afsara Benazir and Felix~Xiaozhu Lin.
\newblock {Safeguarding Privacy in Edge Speech Understanding with Tiny Foundation Models}, 2025.
\newblock URL \url{https://arxiv.org/abs/2502.01649}.

\bibitem[Chen et~al.(2023)Chen, Mo, Wang, Chen, Nie, Wang, and Cui]{chen2023custext}
Sai Chen, Fengran Mo, Yanhao Wang, Cen Chen, Jian-Yun Nie, Chengyu Wang, and Jamie Cui.
\newblock {A Customized Text Sanitization Mechanism with Differential Privacy}.
\newblock In \emph{Findings of the Association for Computational Linguistics: ACL 2023}, pp.\  5747--5758, 2023.
\newblock \doi{10.18653/v1/2023.findings-acl.355}.
\newblock URL \url{https://aclanthology.org/2023.findings-acl.355/}.

\bibitem[{DeepSeek-AI}(2026{\natexlab{a}})]{deepseekai2026deepseekv4}
{DeepSeek-AI}.
\newblock {DeepSeek-V4}: Towards highly efficient million-token context intelligence, 2026{\natexlab{a}}.
\newblock URL \url{https://arxiv.org/abs/2606.19348}.

\bibitem[{DeepSeek-AI}(2026{\natexlab{b}})]{deepseekai2026harness}
{DeepSeek-AI}.
\newblock {DeepSeek Harness}: Everything is a plugin.
\newblock Open-source software repository, 2026{\natexlab{b}}.
\newblock URL \url{https://github.com/deepseek-ai/deepseek-harness}.
\newblock Accessed September 30, 2026.

\bibitem[Dettmers et~al.(2023)Dettmers, Pagnoni, Holtzman, and Zettlemoyer]{dettmers2023qlora}
Tim Dettmers, Artidoro Pagnoni, Ari Holtzman, and Luke Zettlemoyer.
\newblock {QLoRA: Efficient Finetuning of Quantized LLMs}.
\newblock In \emph{Advances in Neural Information Processing Systems}, volume~36, 2023.
\newblock URL \url{https://arxiv.org/abs/2305.14314}.

\bibitem[Duan et~al.(2026)Duan, Zheng, He, Mu, Cai, and Zhang]{duan2026rlaa}
Donghang Duan, Xu~Zheng, Yuefeng He, Chong Mu, Leyi Cai, and Lizong Zhang.
\newblock {Look Twice before You Leap: A Rational Framework for Localized Adversarial Anonymization}.
\newblock In \emph{Findings of the Association for Computational Linguistics: ACL 2026}, 2026.
\newblock URL \url{https://arxiv.org/abs/2512.06713}.

\bibitem[{Eternis}(2025)]{eternis2025anonymizer}
{Eternis}.
\newblock {Anonymizer-4B}.
\newblock Hugging Face model, 2025.
\newblock URL \url{https://huggingface.co/eternisai/Anonymizer-4B}.
\newblock Model card.

\bibitem[Gao et~al.(2023)Gao, Madaan, Zhou, Alon, Liu, Yang, Callan, and Neubig]{gao2023pal}
Luyu Gao, Aman Madaan, Shuyan Zhou, Uri Alon, Pengfei Liu, Yiming Yang, Jamie Callan, and Graham Neubig.
\newblock {PAL: Program-aided Language Models}.
\newblock In \emph{Proceedings of the 40th International Conference on Machine Learning}, volume 202 of \emph{Proceedings of Machine Learning Research}, pp.\  10764--10799, 2023.
\newblock URL \url{https://proceedings.mlr.press/v202/gao23f.html}.

\bibitem[{Gemma Team}(2026)]{gemmateam2026gemma4}
{Gemma Team}.
\newblock {Gemma 4 Technical Report}, 2026.
\newblock URL \url{https://arxiv.org/abs/2607.02770}.

\bibitem[Hartmann et~al.(2024)Hartmann, Tran, Kairouz, C\u{a}rbune, and {Aguera y Arcas}]{hartmann2024help}
Florian Hartmann, Duc-Hieu Tran, Peter Kairouz, Victor C\u{a}rbune, and Blaise {Aguera y Arcas}.
\newblock {Can LLMs get help from other LLMs without revealing private information?}, 2024.
\newblock URL \url{https://arxiv.org/abs/2404.01041}.

\bibitem[Hu et~al.(2026)Hu, Zhu, He, Guo, Zhang, Wang, Guo, Jiang, Hao, Guo, Zhang, Zhang, Yang, Xu, Zhou, and Lin]{hu2026qwen3tts}
Hangrui Hu, Xinfa Zhu, Ting He, Dake Guo, Bin Zhang, Xiong Wang, Zhifang Guo, Ziyue Jiang, Hongkun Hao, Zishan Guo, Xinyu Zhang, Pei Zhang, Baosong Yang, Jin Xu, Jingren Zhou, and Junyang Lin.
\newblock Qwen3-tts technical report.
\newblock \emph{arXiv preprint arXiv:2601.15621}, 2026.

\bibitem[Hui et~al.(2026{\natexlab{a}})Hui, Wu, Qu, Mirmehdi, Rahmani, and Liu]{hui2026visual}
Xiaofei Hui, Qian Wu, Haoxuan Qu, Majid Mirmehdi, Hossein Rahmani, and Jun Liu.
\newblock {When Visual Privacy Protection Meets Multimodal Large Language Models}.
\newblock \emph{International Journal of Computer Vision}, 134\penalty0 (4), 2026{\natexlab{a}}.
\newblock \doi{10.1007/s11263-026-02761-y}.
\newblock URL \url{https://doi.org/10.1007/s11263-026-02761-y}.
\newblock Article 167.

\bibitem[Hui et~al.(2026{\natexlab{b}})Hui, Dong, Sivapiromrat, Shareghi, and Collier]{hui2026privacyr1}
Zheng Hui, Yijiang~River Dong, Sanhanat Sivapiromrat, Ehsan Shareghi, and Nigel Collier.
\newblock {Privacy-R1: Privacy-Aware Multi-LLM Agent Collaboration via Reinforcement Learning}.
\newblock In \emph{Proceedings of the 64th Annual Meeting of the Association for Computational Linguistics}, 2026{\natexlab{b}}.
\newblock URL \url{https://arxiv.org/abs/2510.16054}.

\bibitem[Hunter(2007)]{hunter2007matplotlib}
John~D. Hunter.
\newblock Matplotlib: A 2d graphics environment.
\newblock \emph{Computing in Science \& Engineering}, 9\penalty0 (3):\penalty0 90--95, 2007.
\newblock \doi{10.1109/MCSE.2007.55}.

\bibitem[Jin et~al.(2025)Jin, Cao, Su, Wang, Zhang, Xue, Hao, Dong, and Yang]{jin2025whispering}
Weifei Jin, Yuxin Cao, Junjie Su, Derui Wang, Yedi Zhang, Minhui Xue, Jie Hao, Jin~Song Dong, and Yixian Yang.
\newblock {Whispering Under the Eaves: Protecting User Privacy Against Commercial and {LLM-powered} Automatic Speech Recognition Systems}.
\newblock In \emph{34th USENIX Security Symposium (USENIX Security 25)}, pp.\  4643--4662, Seattle, WA, August 2025. USENIX Association.
\newblock ISBN 978-1-939133-52-6.
\newblock URL \url{https://www.usenix.org/conference/usenixsecurity25/presentation/jin-weifei}.

\bibitem[Lin et~al.(2022)Lin, Chuang, Chung, Yang, Chen, Dong, Li, Mohamed, Lee, and Lee]{lin2022dual}
Guan-Ting Lin, Yung-Sung Chuang, Ho-Lam Chung, Shu-wen Yang, Hsuan-Jui Chen, Shuyan~Annie Dong, Shang-Wen Li, Abdelrahman Mohamed, Hung-yi Lee, and Lin-shan Lee.
\newblock {DUAL: Discrete Spoken Unit Adaptive Learning for Textless Spoken Question Answering}.
\newblock In \emph{Interspeech 2022}, pp.\  5165--5169, 2022.
\newblock \doi{10.21437/Interspeech.2022-612}.
\newblock URL \url{https://www.isca-archive.org/interspeech_2022/lin22c_interspeech.html}.

\bibitem[Liu et~al.(2023)Liu, Eisenschlos, Piccinno, Krichene, Pang, Lee, Joshi, Chen, Collier, and Altun]{liu2023deplot}
Fangyu Liu, Julian Eisenschlos, Francesco Piccinno, Syrine Krichene, Chenxi Pang, Kenton Lee, Mandar Joshi, Wenhu Chen, Nigel Collier, and Yasemin Altun.
\newblock {DePlot: One-shot visual language reasoning by plot-to-table translation}.
\newblock In \emph{Findings of the Association for Computational Linguistics: ACL 2023}, pp.\  10381--10399, 2023.
\newblock \doi{10.18653/v1/2023.findings-acl.660}.
\newblock URL \url{https://aclanthology.org/2023.findings-acl.660/}.

\bibitem[Liu et~al.(2026)Liu, Xie, Yu, Zhu, Li, and Chen]{liu2026veilmind}
Zhenhua Liu, Zhanxu Xie, Junjie Yu, Tong Zhu, Lijun Li, and Wenliang Chen.
\newblock {Demystifying the Privacy-Utility Trade-off in LLM Interactions}, 2026.
\newblock URL \url{https://arxiv.org/abs/2609.10992}.

\bibitem[Masry et~al.(2022)Masry, Long, Tan, Joty, and Hoque]{masry2022chartqa}
Ahmed Masry, Do~Xuan Long, Jia~Qing Tan, Shafiq Joty, and Enamul Hoque.
\newblock {ChartQA: A Benchmark for Question Answering about Charts with Visual and Logical Reasoning}.
\newblock In \emph{Findings of the Association for Computational Linguistics: ACL 2022}, pp.\  2263--2279, 2022.
\newblock \doi{10.18653/v1/2022.findings-acl.177}.
\newblock URL \url{https://aclanthology.org/2022.findings-acl.177/}.

\bibitem[Meisenbacher et~al.(2024)Meisenbacher, Chevli, Vladika, and Matthes]{meisenbacher2024dpmlm}
Stephen Meisenbacher, Maulik Chevli, Juraj Vladika, and Florian Matthes.
\newblock {DP-MLM: Differentially Private Text Rewriting Using Masked Language Models}.
\newblock In \emph{Findings of the Association for Computational Linguistics: ACL 2024}, pp.\  9314--9328, 2024.
\newblock \doi{10.18653/v1/2024.findings-acl.554}.
\newblock URL \url{https://aclanthology.org/2024.findings-acl.554/}.

\bibitem[Meisenbacher et~al.(2025)Meisenbacher, Chevli, and Matthes]{meisenbacher2025dpst}
Stephen Meisenbacher, Maulik Chevli, and Florian Matthes.
\newblock {Leveraging Semantic Triples for Private Document Generation with Local Differential Privacy Guarantees}.
\newblock In \emph{Proceedings of the 2025 Conference on Empirical Methods in Natural Language Processing}, pp.\  8987--9003, 2025.
\newblock URL \url{https://arxiv.org/abs/2508.20736}.

\bibitem[Meyer et~al.(2022)Meyer, Lux, Denisov, Koch, Tilli, and Vu]{meyer2022anonymization}
Sarina Meyer, Florian Lux, Pavel Denisov, Julia Koch, Pascal Tilli, and Ngoc~Thang Vu.
\newblock {Speaker Anonymization with Phonetic Intermediate Representations}.
\newblock In \emph{Interspeech 2022}, pp.\  4925--4929, 2022.
\newblock \doi{10.21437/Interspeech.2022-10703}.
\newblock URL \url{https://www.isca-archive.org/interspeech_2022/meyer22b_interspeech.html}.

\bibitem[Mishra et~al.(2025)Mishra, Li, Fu, Noh, and Kim]{mishra2025revision}
Abhijit Mishra, Mingda Li, Hsiang Fu, Richard Noh, and Minji Kim.
\newblock {ReVision: A Dataset and Baseline VLM for Privacy-Preserving Task-Oriented Visual Instruction Rewriting}.
\newblock In \emph{Proceedings of the 14th International Joint Conference on Natural Language Processing and the 4th Conference of the Asia-Pacific Chapter of the Association for Computational Linguistics}, pp.\  2877--2889, 2025.
\newblock \doi{10.18653/v1/2025.ijcnlp-long.154}.
\newblock URL \url{https://aclanthology.org/2025.ijcnlp-long.154/}.

\bibitem[{OpenAI}(2026{\natexlab{a}})]{openai2026codexcli}
{OpenAI}.
\newblock {Codex CLI}.
\newblock Open-source software repository, 2026{\natexlab{a}}.
\newblock URL \url{https://github.com/openai/codex}.
\newblock Accessed September 30, 2026.

\bibitem[{OpenAI}(2026{\natexlab{b}})]{openai2026gpt56}
{OpenAI}.
\newblock {GPT-5.6 System Card}, July 2026{\natexlab{b}}.
\newblock URL \url{https://deploymentsafety.openai.com/gpt-5-6}.
\newblock System card for the GPT-5.6 family, including GPT-5.6 Sol.

\bibitem[{Qwen Team}(2026{\natexlab{a}})]{qwenteam2026qwen35omni}
{Qwen Team}.
\newblock {Qwen3.5-Omni Technical Report}, 2026{\natexlab{a}}.
\newblock URL \url{https://arxiv.org/abs/2604.15804}.

\bibitem[{Qwen Team}(2026{\natexlab{b}})]{qwenteam2026qwen37}
{Qwen Team}.
\newblock {Qwen3.7}: The agent frontier, May 2026{\natexlab{b}}.
\newblock URL \url{https://qwen.ai/blog?id=qwen3.7}.

\bibitem[Shen et~al.(2025)Shen, Xu, Yin, and Yang]{shen2025psic}
Xuelin Shen, Jiayin Xu, Kangsheng Yin, and Wenhan Yang.
\newblock Privacy-shielded image compression: Defending against exploitation from vision-language pretrained models.
\newblock In \emph{Proceedings of the 42nd International Conference on Machine Learning}, volume 267 of \emph{Proceedings of Machine Learning Research}, pp.\  54600--54610. PMLR, 2025.
\newblock URL \url{https://proceedings.mlr.press/v267/shen25k.html}.

\bibitem[Shi et~al.(2026)Shi, Wang, Guo, Wang, Zhang, Zhang, Guo, Hao, Xi, Yang, Xu, Zhou, and Lin]{shi2026qwen3asr}
Xian Shi, Xiong Wang, Zhifang Guo, Yongqi Wang, Pei Zhang, Xinyu Zhang, Zishan Guo, Hongkun Hao, Yu~Xi, Baosong Yang, Jin Xu, Jingren Zhou, and Junyang Lin.
\newblock {Qwen3-ASR Technical Report}, 2026.
\newblock URL \url{https://arxiv.org/abs/2601.21337}.

\bibitem[Siyan et~al.(2025)Siyan, Raghuram, Khattab, Hirschberg, and Yu]{siyan2025papillon}
Li~Siyan, Vethavikashini~Chithrra Raghuram, Omar Khattab, Julia Hirschberg, and Zhou Yu.
\newblock {PAPILLON: Privacy Preservation from Internet-based and Local Language Model Ensembles}.
\newblock In \emph{Proceedings of the 2025 Conference of the Nations of the Americas Chapter of the Association for Computational Linguistics: Human Language Technologies (Volume 1: Long Papers)}, pp.\  3371--3390, 2025.
\newblock \doi{10.18653/v1/2025.naacl-long.173}.
\newblock URL \url{https://aclanthology.org/2025.naacl-long.173/}.

\bibitem[Siyan et~al.(2026)Siyan, Yu, and Hirschberg]{siyan2026beyonddirect}
Li~Siyan, Zhou Yu, and Julia Hirschberg.
\newblock {Beyond Direct Identifiers: Probabilistic Privacy Risk Estimation for Privacy-Conscious LLM Query Delegation}.
\newblock In \emph{HAIPS Workshop at COLM 2026}, 2026.
\newblock URL \url{https://arxiv.org/abs/2608.09140}.

\bibitem[Staab et~al.(2024)Staab, Vero, Balunovi\'{c}, and Vechev]{staab2024beyond}
Robin Staab, Mark Vero, Mislav Balunovi\'{c}, and Martin Vechev.
\newblock {Beyond Memorization: Violating Privacy via Inference with Large Language Models}.
\newblock In \emph{The Twelfth International Conference on Learning Representations}, 2024.
\newblock URL \url{https://openreview.net/forum?id=kmn0BhQk7p}.

\bibitem[Staab et~al.(2025)Staab, Vero, Balunovi\'{c}, and Vechev]{staab2025anonymizers}
Robin Staab, Mark Vero, Mislav Balunovi\'{c}, and Martin Vechev.
\newblock {Language Models are Advanced Anonymizers}.
\newblock In \emph{The Thirteenth International Conference on Learning Representations}, 2025.
\newblock URL \url{https://openreview.net/forum?id=82p8VHRsaK}.

\bibitem[Thareja et~al.(2025)Thareja, Nakov, Vepakomma, and Lukas]{thareja2025dpfusion}
Rushil Thareja, Preslav Nakov, Praneeth Vepakomma, and Nils Lukas.
\newblock {DP-Fusion: Token-Level Differentially Private Inference for Large Language Models}, 2025.
\newblock URL \url{https://arxiv.org/abs/2507.04531}.

\bibitem[Tian et~al.(2026)Tian, Bhattacharjee, Hanson, Rubin, Lo, and Tandon]{tian2026stamp}
Fengwei Tian, Payel Bhattacharjee, Heidi Hanson, Geoffrey~D. Rubin, Joseph~Y. Lo, and Ravi Tandon.
\newblock {STAMP: Selective Task-Aware Mechanism for Text Privacy}.
\newblock In \emph{Proceedings of the 19th Conference of the European Chapter of the Association for Computational Linguistics (Volume 1: Long Papers)}, pp.\  1319--1333, 2026.
\newblock \doi{10.18653/v1/2026.eacl-long.61}.
\newblock URL \url{https://aclanthology.org/2026.eacl-long.61/}.

\bibitem[T\"{o}mek\c{c}e et~al.(2024)T\"{o}mek\c{c}e, Vero, Staab, and Vechev]{tomekce2024attributes}
Batuhan T\"{o}mek\c{c}e, Mark Vero, Robin Staab, and Martin Vechev.
\newblock {Private Attribute Inference from Images with Vision-Language Models}.
\newblock In \emph{Advances in Neural Information Processing Systems}, volume~37, pp.\  103619--103651, 2024.
\newblock URL \url{https://proceedings.neurips.cc/paper_files/paper/2024/hash/bb97e9a7c811904c9b01f51fde66edcf-Abstract-Conference.html}.

\bibitem[Tong et~al.(2025)Tong, Chen, Zhang, Qi, Zhang, Yu, Zhang, and Zhang]{tong2025inferdpt}
Meng Tong, Kejiang Chen, Jie Zhang, Yuang Qi, Weiming Zhang, Nenghai Yu, Tianwei Zhang, and Zhikun Zhang.
\newblock {InferDPT: Privacy-Preserving Inference for Closed-Box Large Language Models}.
\newblock \emph{IEEE Transactions on Dependable and Secure Computing}, 22\penalty0 (5):\penalty0 4625--4640, 2025.
\newblock URL \url{https://ieeexplore.ieee.org/document/10922117/}.

\bibitem[Utpala et~al.(2023)Utpala, Hooker, and Chen]{utpala2023dpprompt}
Saiteja Utpala, Sara Hooker, and Pin-Yu Chen.
\newblock {Locally Differentially Private Document Generation Using Zero Shot Prompting}.
\newblock In \emph{Findings of the Association for Computational Linguistics: EMNLP 2023}, pp.\  8442--8457, 2023.
\newblock \doi{10.18653/v1/2023.findings-emnlp.566}.
\newblock URL \url{https://aclanthology.org/2023.findings-emnlp.566/}.

\bibitem[Wang et~al.(2025)Wang, Lin, Yu, and Zhang]{wang2025spti}
Haoxiang Wang, Zinan Lin, Da~Yu, and Huishuai Zhang.
\newblock {Synthesize Privacy-Preserving High-Resolution Images via Private Textual Intermediaries}.
\newblock In \emph{Advances in Neural Information Processing Systems}, volume~38, 2025.
\newblock URL \url{https://arxiv.org/abs/2506.07555}.

\bibitem[Wei et~al.(2025)Wei, Wang, jae Kim, and Chen]{wei2025spokenmathematicalreasoningbenchmarking}
Chengwei Wei, Bin Wang, Jung jae Kim, and Nancy~F. Chen.
\newblock Towards spoken mathematical reasoning: Benchmarking speech-based models over multi-faceted math problems, 2025.
\newblock URL \url{https://arxiv.org/abs/2505.15000}.

\bibitem[Xia et~al.(2025)Xia, Zhang, Ye, Yan, Liu, Zhou, Chen, Ye, Dou, Shi, Yan, and Qiao]{xia2025chartxchartvlmversatile}
Renqiu Xia, Bo~Zhang, Hancheng Ye, Xiangchao Yan, Qi~Liu, Hongbin Zhou, Zijun Chen, Peng Ye, Min Dou, Botian Shi, Junchi Yan, and Yu~Qiao.
\newblock Chartx \& chartvlm: A versatile benchmark and foundation model for complicated chart reasoning, 2025.
\newblock URL \url{https://arxiv.org/abs/2402.12185}.

\bibitem[Xu et~al.(2025)Xu, Guo, He, Hu, He, Bai, Chen, Wang, Fan, Dang, Zhang, Wang, Chu, and Lin]{xu2025qwen25omni}
Jin Xu, Zhifang Guo, Jinzheng He, Hangrui Hu, Ting He, Shuai Bai, Keqin Chen, Jialin Wang, Yang Fan, Kai Dang, Bin Zhang, Xiong Wang, Yunfei Chu, and Junyang Lin.
\newblock {Qwen2.5-Omni Technical Report}, 2025.
\newblock URL \url{https://arxiv.org/abs/2503.20215}.

\bibitem[Yang et~al.(2025)Yang, Li, Yang, Zhang, Hui, Zheng, Yu, Gao, Huang, Lv, Zheng, Liu, Zhou, Huang, Hu, Ge, Wei, Lin, Tang, Yang, Tu, Zhang, Yang, Yang, Zhou, Zhou, Lin, Dang, Bao, Yang, Yu, Deng, Li, Xue, Li, Zhang, Wang, Zhu, Men, Gao, Liu, Luo, Li, Tang, Yin, Ren, Wang, Zhang, Ren, Fan, Su, Zhang, Zhang, Wan, Liu, Wang, Cui, Zhang, Zhou, and Qiu]{yang2025qwen3}
An~Yang, Anfeng Li, Baosong Yang, Beichen Zhang, Binyuan Hui, Bo~Zheng, Bowen Yu, Chang Gao, Chengen Huang, Chenxu Lv, Chujie Zheng, Dayiheng Liu, Fan Zhou, Fei Huang, Feng Hu, Hao Ge, Haoran Wei, Huan Lin, Jialong Tang, Jian Yang, Jianhong Tu, Jianwei Zhang, Jianxin Yang, Jiaxi Yang, Jing Zhou, Jingren Zhou, Junyang Lin, Kai Dang, Keqin Bao, Kexin Yang, Le~Yu, Lianghao Deng, Mei Li, Mingfeng Xue, Mingze Li, Pei Zhang, Peng Wang, Qin Zhu, Rui Men, Ruize Gao, Shixuan Liu, Shuang Luo, Tianhao Li, Tianyi Tang, Wenbiao Yin, Xingzhang Ren, Xinyu Wang, Xinyu Zhang, Xuancheng Ren, Yang Fan, Yang Su, Yichang Zhang, Yinger Zhang, Yu~Wan, Yuqiong Liu, Zekun Wang, Zeyu Cui, Zhenru Zhang, Zhipeng Zhou, and Zihan Qiu.
\newblock {Qwen3 Technical Report}.
\newblock \emph{arXiv preprint arXiv:2505.09388}, 2025.
\newblock URL \url{https://arxiv.org/abs/2505.09388}.

\bibitem[Yue et~al.(2021)Yue, Du, Wang, Li, Sun, and Chow]{yue2021santext}
Xiang Yue, Minxin Du, Tianhao Wang, Yaliang Li, Huan Sun, and Sherman S.~M. Chow.
\newblock {Differential Privacy for Text Analytics via Natural Text Sanitization}.
\newblock In \emph{Findings of the Association for Computational Linguistics: ACL-IJCNLP 2021}, pp.\  3853--3866, 2021.
\newblock \doi{10.18653/v1/2021.findings-acl.337}.
\newblock URL \url{https://aclanthology.org/2021.findings-acl.337/}.

\bibitem[Zhang et~al.(2024)Zhang, Wang, Hua, Qi, Ding, and Zhou]{zhang2024cogenesis}
Kaiyan Zhang, Jianyu Wang, Ermo Hua, Biqing Qi, Ning Ding, and Bowen Zhou.
\newblock {CoGenesis: A Framework Collaborating Large and Small Language Models for Secure Context-Aware Instruction Following}.
\newblock In \emph{Proceedings of the 62nd Annual Meeting of the Association for Computational Linguistics}, 2024.
\newblock URL \url{https://arxiv.org/abs/2403.03129}.

\bibitem[Zhang et~al.(2026)Zhang, Lu, Zhou, Xu, Cheong, Lu, and Wang]{zhang2026collaborative}
Min Zhang, Yuzhe Lu, Yun Zhou, Panpan Xu, Lin~Lee Cheong, Chang-Tien Lu, and Haozhu Wang.
\newblock {Collaborative LLM Numerical Reasoning with Local Data Protection}.
\newblock In \emph{Proceedings of the AAAI Conference on Artificial Intelligence}, volume~40, 2026.
\newblock \doi{10.1609/aaai.v40i41.40779}.
\newblock URL \url{https://ojs.aaai.org/index.php/AAAI/article/view/40779}.

\end{thebibliography}

\appendix
\begingroup
\setlength{\parskip}{2pt}
\setlength{\parsep}{0pt}
\setlength{\topsep}{2pt}
\setlength{\itemsep}{1pt}
\setlength{\intextsep}{3pt}
\setlength{\textfloatsep}{3pt}
\setlength{\floatsep}{3pt}
\setlength{\abovecaptionskip}{2pt}
\setlength{\belowcaptionskip}{0pt}
\renewcommand{\arraystretch}{1.04}
\raggedbottom
\section{Implementation Details}
\label{app:implementation}

\subsection{Evaluation Data}
\label{app:eval-data}

\begin{table}[H]
\centering
\small
\setlength{\tabcolsep}{5pt}
\caption{Construction of the evaluation subsets. ID denotes
in-distribution evaluation, while OOD denotes datasets not used for
Rewriter distillation or checkpoint selection. All evaluation sample
IDs are fixed before model execution.}
\vspace{2pt}
\label{tab:evaluation-subsets}

\begin{tabularx}{\linewidth}{
@{}l
>{\raggedright\arraybackslash}X
>{\raggedright\arraybackslash}X
>{\raggedright\arraybackslash}X
>{\raggedright\arraybackslash}X@{}
}
\toprule
\textbf{Item}
& \textbf{ChartQA}
& \textbf{NMSQA}
& \textbf{ChartX}
& \textbf{Spoken-MQA} \\
\midrule

Evaluation setting
& ID
& ID
& OOD
& OOD \\

Evaluated examples
& 2,000
& 2,000
& 2,000
& 2,000 \\

Source pool
& Held-out evaluation partition
& Held-out evaluation partition
& Official benchmark pool
& Official benchmark pool \\

Evidence modality
& Image
& Audio
& Image
& Audio \\

Query / instruction modality
& Text
& Text
& Text
& Audio problem with textual instruction \\

Eligibility
& Single numeric answer; readable PNG; duplicate questions removed
& Single consistent numeric reference; passage length $\leq 350$ words
& QA task; parseable single numeric answer; readable chart image
& Valid audio; parseable numeric reference \\

Rewriter distillation overlap
& None
& None
& None
& None \\

Selection
& Fixed before execution
& Fixed before execution
& Fixed before execution
& Fixed before execution \\

\bottomrule
\end{tabularx}
\end{table}

\paragraph{ID evaluation subsets.}
We split the selected ChartQA~\citep{masry2022chartqa} and
NMSQA~\citep{lin2022dual} records into disjoint training and test
partitions and evaluate 2,000 eligible numeric-answer examples from each
test partition. Eligible records are shuffled with seed 42 after removing duplicate
media paths, and the selected sample IDs are fixed before evaluation. Distillation and
checkpoint selection use only training-side records.

\paragraph{OOD evaluation subsets.}
For out-of-distribution evaluation, we independently sample 2,000
numeric-answer examples from ChartX and 2,000 from Spoken-MQA.
Neither dataset contributes records to Rewriter distillation or
checkpoint selection. The selected sample IDs are fixed before
evaluation and recorded in the run manifests.

\subsection{Components}
\label{app:components}

The implementation uses Qwen3-VL~\citep{bai2025qwen3vl} for chart capture,
Qwen3-ASR~\citep{shi2026qwen3asr} for speech capture,
Qwen3-4B~\citep{yang2025qwen3} with QLoRA~\citep{dettmers2023qlora}
for rewriting, DeepSeek-V4~\citep{deepseekai2026deepseekv4} for chart
specifications, and Qwen3-TTS~\citep{hu2026qwen3tts} for synthesis.
The remote solver is Qwen3.5-Omni-Plus~\citep{qwenteam2026qwen35omni};
the local references include Gemma 4~\citep{gemmateam2026gemma4} and
Qwen2.5-Omni~\citep{xu2025qwen25omni}. Codex CLI~\citep{openai2026codexcli} and DeepSeek Harness (DSH)~\citep{deepseekai2026harness} are alternative Stage~3
implementations that receive the same protected records as \textsc{ReCast} and
reconstruct the target modality before evaluation by the same remote solver.

\begin{table}[!t]
\centering
\caption{Component configuration.}
\label{tab:a-2}
\small
\renewcommand{\arraystretch}{1.04}
\setlength{\tabcolsep}{4pt}

\begin{tabularx}{\linewidth}{
    >{\raggedright\arraybackslash}p{0.17\linewidth}
    >{\raggedright\arraybackslash}X
    >{\raggedright\arraybackslash}p{0.14\linewidth}
    >{\raggedright\arraybackslash}p{0.22\linewidth}
}
\toprule
Component & Model or implementation & Location & Input \\
\midrule
Text parser & deterministic parser with field and schema validation & Local & Original text evidence and question \\
Chart-to-text & Qwen3-VL-8B-Instruct, question-aware prompt & Local & Original chart and question \\
Speech-to-text & Qwen3-ASR-1.7B & Local & Evidence audio \\
Numeral normalization & \texttt{text2num}, isolated number words included & Local & Evidence, question, choices \\
Rewriter & Qwen3-4B + distilled QLoRA adapter & Local & Normalized five-field record \\
Numerical map & role-aware banded map (Appendix~\ref{app:numerical-map}) & Local & Rewritten evidence and question \\
Distillation teacher & GPT-5.6-Sol~\citep{openai2026gpt56} & Remote, offline & Training-side records \\
ChartSpec generation & DeepSeek \texttt{deepseek-v4-flash}, JSON output & Remote & Protected evidence and question \\
ChartSpec review & \texttt{deepseek-v4-flash-vision-exp} & Remote & Specification, protected text, rendered chart \\
Chart rendering & fixed Matplotlib templates (bar, grouped bar, stacked bar, line, pie, table) & Local & Validated ChartSpec \\
Speech synthesis & Qwen3-TTS-Flash & Remote & Protected passage \\
Solver & Qwen3.5-Omni-Plus (DashScope compatible mode) & Remote & Protected artifact and question \\
Program execution & separate process, 5 s timeout & Local & Restored program \\
Remote-native reference & Qwen3.5-Omni-Plus, same solver prompt, no inverse map & Remote & Original artifact and question \\
Local references & Gemma-4-E4B; Qwen2.5-Omni-3B; Qwen3-VL-8B-Instruct image-to-code (ChartQA) & Local & Original artifact and question \\
Leakage judge & \texttt{qwen3.5-omni-plus-2026-03-15}, native multimodal & Remote, evaluation only & Original and solver-bound contexts \\
\bottomrule
\end{tabularx}
\end{table}

Algorithm~\ref{alg:ModalGuard} summarizes the transformation of a single
request, with $r=m$ corresponding to same-modality release.

\begin{algorithm}[t]
\caption{\textsc{ReCast} transformation for one request}
\label{alg:ModalGuard}
\begin{algorithmic}[1]

\Require Private source $X$, query $q$, optional choices $C$,
input modality $m$, output modality $r$
\Ensure Protected pair $(\widetilde{X},\widetilde{q})$ and locally retained inverse map $h^{-1}$, or failure

\Statex \textbf{Stage 1: Evidence Capture}

\If{$m=\mathrm{text}$}
    \State $c \gets \Call{ParseText}{X}$
\ElsIf{$m=\mathrm{image}$}
    \State $c \gets \Call{ChartToText}{X,q}$
\ElsIf{$m=\mathrm{audio}$}
    \State $c \gets \Call{ASR}{X}$
\EndIf

\State $\mathcal{R} \gets
\{\mathrm{id},\,\mathrm{modality}{:}m,\,
\mathrm{context}{:}c,\,
\mathrm{question}{:}q,\,
\mathrm{choices}{:}C\}$

\State $\mathcal{R}_{\mathrm{num}}
    \gets \Call{NormalizeNumerals}{\mathcal{R}}$
\State $D \gets \mathcal{R}_{\mathrm{num}}$
    \Comment{Retained locally for source-consistency checking}

\Statex \textbf{Stage 2: Joint Rewriting and Numerical Protection}

\Repeat
    \State $\overline{\mathcal{R}}
        \gets \Call{Rewriter}{\mathcal{R}_{\mathrm{num}}}$
\Until{$\Call{Validate}{\overline{\mathcal{R}}}$}

\State $h \gets
    \Call{BuildRoleAwareMap}
    {\overline{\mathcal{R}}.\mathrm{context},
     \overline{\mathcal{R}}.\mathrm{question}}$

\State $(\widetilde{c},\widetilde{q})
    \gets
    \Call{ApplyMap}
    {h,\overline{\mathcal{R}}.\mathrm{context},
     \overline{\mathcal{R}}.\mathrm{question}}$

\Statex \textbf{Stage 3: Modality Reconstruction and Validation}

\If{$r=\mathrm{text}$}
    \State $\widetilde{X}
        \gets \Call{SerializeProtectedText}{\widetilde{c}}$

\ElsIf{$r=\mathrm{image}$}
    \State $\widetilde{X}
        \gets \Call{SpecRenderReviewRepair}{\widetilde{c},\widetilde{q}}$

\ElsIf{$r=\mathrm{audio}$}
    \State $\widetilde{X}
        \gets \Call{SynthesizeAndCheckSpeech}{\widetilde{c},\widetilde{q}}$
\EndIf

\If{$\neg\Call{ReleaseCheck}
    {X,q,\widetilde{D},\widetilde{X},\widetilde{q},h^{-1}}$}
    \State \Return \textsc{ReleaseFailure}
\EndIf

\State \Return $(\widetilde{X},\widetilde{q})$
\Comment{Keep $h^{-1}$ local}

\end{algorithmic}
\end{algorithm}

\subsection{Numerical Mapping and Recovery}
\label{app:numerical-map}

Currency prefixes and thousands separators are normalized, so that each repeated value has one identity. Nonnegative integers from 0 to 15 remain unchanged unless detected as years.
These small integers are intentionally preserved because they commonly encode
task-structural quantities such as counts or indices rather than
source-specific numerical content.
Years are identified from context, as in ``in 2016'', ``from 2014 to 2017'', or ``fiscal year 2015'', and grouped by era. Each group receives a shared random offset, which preserves differences within the group. The remaining values fall into three magnitude bands: below 1,263, from 1,263 to 2,100, and above 2,100. Within each band, random two-decimal replacements are drawn from a target range and sorted to preserve order, with checks against collisions with other replacements and preserved integers. For presentation, trailing zeros are omitted, so a sampled value such
as 48.00 may be displayed as 48. Each original value receives the same replacement throughout evidence and query. Deployment draws a fresh map for each request. For reproducibility,
the mapping RNG is initialized once with seed 42 at the beginning of
each experimental run and is not reset for individual requests.

The solver marks observed operands as \texttt{protected(<number>)}. Recovery maps each marked literal through $h^{-1}$ and leaves preserved integers unchanged. The restored program then computes on the original values.

\subsection{Distillation Data}
\label{app:distillation}

Each teacher request rewrites one five-field record, with a fixed two-example demonstration (one chart, one audio). An output is accepted if it satisfies all of the following:
\begin{itemize}
    \item it is a one-element JSON array with exactly the five fields;
    \item \texttt{id} and \texttt{modality} are byte-identical to the input;
    \item \texttt{choices} keep their length and order;
    \item \texttt{context} and \texttt{question} are nonempty;
    \item every field keeps its sequence of numerical literals.
\end{itemize}

Accepted pairs are deterministically split into training and development sets.

\begin{table}[H]
\centering
\small
\setlength{\tabcolsep}{3pt}
\caption{Distillation corpus.}
\label{tab:a-3}
\begin{tabular}{lrrr}
\toprule
Item & Total & NMSQA & ChartQA \\
\midrule
Source records & 16,370 & 7,000 & 9,370 \\
Accepted pairs & 15,795 & 6,742 & 9,053 \\
Rejected (changed numerical literals) & 575 & 258 & 317 \\
Training pairs & 15,480 & 6,604 & 8,876 \\
Validation pairs & 315 & 138 & 177 \\
\bottomrule
\end{tabular}
\end{table}

Of the accepted pairs, 129 are reused from an earlier accepted pass, and 1,718 change the number of sentences, which the filters allow. We release the accepted teacher traces, so the student can be retrained directly from them.

Figs.~\ref{fig:distill-types}-\ref{fig:distill-wordcloud} characterize the 15,795 accepted pairs. The chart records cover all four chart types with a consistent mix of question operations (Figs.~\ref{fig:distill-types} and~\ref{fig:distill-heatmap}), and both modalities span a broad range of topics (Fig.~\ref{fig:distill-topics}). Rewriting replaces the source vocabulary with substitute entities and topics (Fig.~\ref{fig:distill-wordcloud}). Chart types, question operations, and topics are assigned by keyword rules.

\begingroup
\setlength{\intextsep}{6pt}
\setlength{\textfloatsep}{6pt}
\setlength{\floatsep}{6pt}

\begin{figure}[H]
\centering
\includegraphics[
    width=\linewidth,
    trim={0mm 0mm 0mm 8mm},
    clip
]{\detokenize{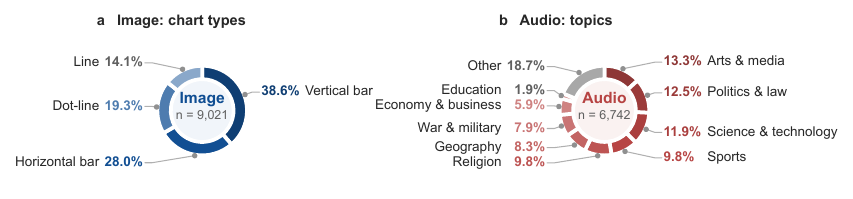}}
\caption{Chart types of the image records (a; 32 records with no identifiable type are omitted) and topics of the audio records (b) in the distillation corpus.}
\label{fig:distill-types}
\end{figure}

\begin{figure}[H]
\centering
\includegraphics[
    width=0.95\linewidth,
    trim={0mm 0mm 0mm 8mm},
    clip
]{\detokenize{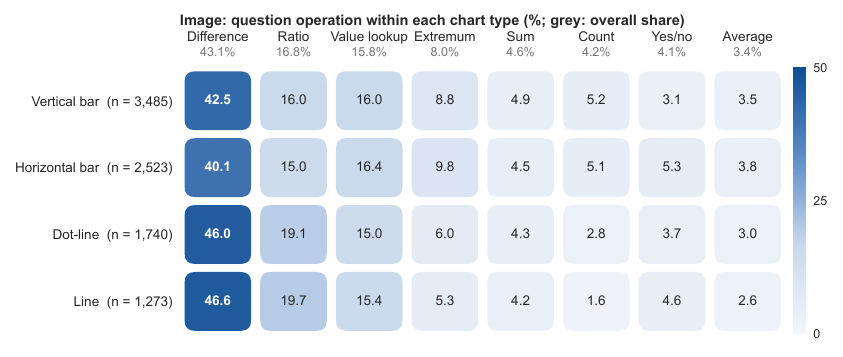}}
\caption{Question operations within each chart type of the image records. Each row sums to 100\%; grey numbers give the share of each operation over all image records.}
\label{fig:distill-heatmap}
\end{figure}

\begin{figure}[H]
\centering
\includegraphics[
    width=\linewidth
]{\detokenize{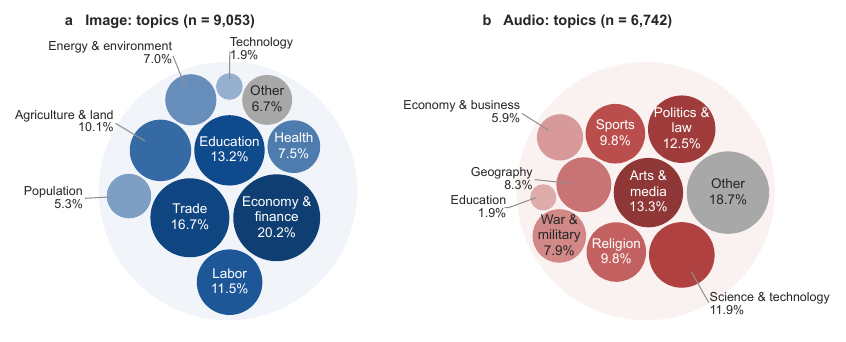}}
\caption{Topics of the image (a) and audio (b) records. Circle area is proportional to the share of records; circle position carries no meaning.}
\label{fig:distill-topics}
\end{figure}

\begin{figure}[t]
\centering
\includegraphics[
    width=\linewidth
]{\detokenize{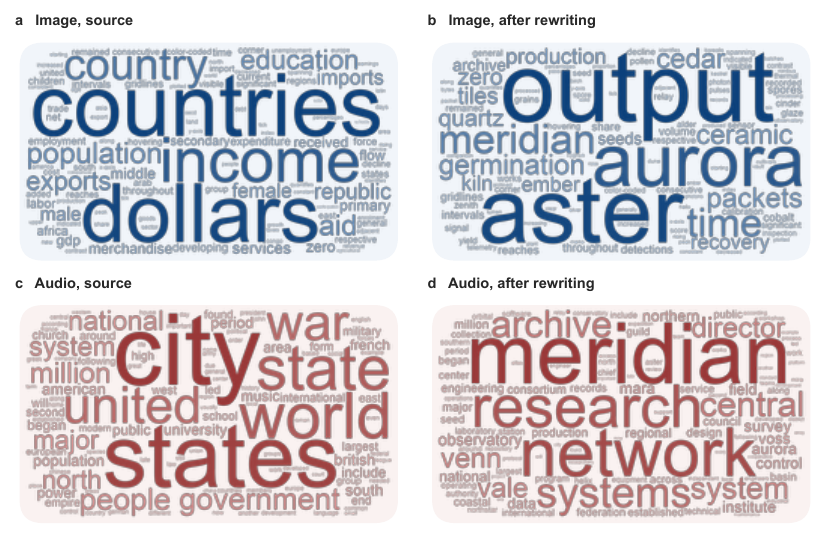}}
\caption{Most frequent content words in the image (a, b) and audio (c, d) records before and after teacher rewriting. Word size and shade increase with frequency; chart-description and function words are removed.}
\label{fig:distill-wordcloud}
\end{figure}

\endgroup

\subsection{Request Logs}
\label{app:request-logs}

Each solver request is logged before transmission with its messages, model identifier, decoding settings, endpoint, system prompt, and attached media; logging must succeed before the network call proceeds. Transport retries and format repairs receive distinct event identifiers with parent links, and the leakage audit aggregates all solver events of an example. ChartSpec generation, chart review, and TTS requests are logged as a separate reconstruction boundary containing the protected record, specification candidates, review inputs, TTS passage, and response metadata. Original inputs and inverse maps stay local; the leakage judge receives original inputs only inside the evaluation workflow.

\begingroup
\let\FloatBarrier\relax
\section{Evaluation Details}
\label{app:evaluation}
\endgroup

\subsection{Answer Matching}
\label{app:answer-matching}

For numerical prediction $p$ and acceptable reference $g$, a prediction is correct if

\begin{equation}
    |p-g|\leq\max\!\left(10^{-6},\ r_d\max(|p|,|g|,1)\right),
\end{equation}

with $r_d=10^{-6}$ for ChartQA and NMSQA. Nonnumeric aliases use normalized exact matching. Acceptable references are derived from the original answer, question, and trusted local evidence, never from predictions. For ChartQA, an unambiguous million or billion unit in the local chart description admits both the scaled and absolute value, and questions asking for a percentage, rate, ratio, or probability admit both proportion and percentage-point forms. Years, structural counts, and denominator units such as ``per million'' are not expanded. The evaluator requires identical sample IDs across compared methods.

\subsection{Leakage Aggregation}
\label{app:leakage-aggregation}

For image inputs, the original context $A_i$ contains the image and query. For audio inputs, it contains the audio, query, and trusted local transcript. The solver-bound context $B_i$ contains all recorded solver requests of example $i$. The judge returns four binary labels for each pair $(A_i, B_i)$, and the example counts as leaked if any label is positive. In the reported experiments, every sample ID eventually produced a
solver-bound request after validation; no sample ended in a terminal
release failure. Therefore, the leakage rate over all examples equals
the rate over delivered requests. All \textsc{ReCast} rows of Tabs.~\ref{tab:1} and~\ref{tab:3} have resolved labels for every example.

\subsection{Local References}
\label{app:local-references}

The evaluated local models are Gemma 4~\citep{gemmateam2026gemma4},
Qwen2.5-Omni~\citep{xu2025qwen25omni}, and
Qwen3-VL~\citep{bai2025qwen3vl}.

\begin{table}[!t]
\centering\small
\setlength{\tabcolsep}{3pt}
\caption{Accuracy of evaluated local references.}
\label{tab:b-1}
\begin{tabular}{lrrr}
\toprule
Model & ChartQA & NMSQA & Overall \\
\midrule
Gemma-4-E4B & 49.50 & 44.60 & 47.05 \\
Qwen2.5-Omni-3B & 23.70 & 46.30 & 35.00 \\
Qwen3-VL-8B-Instruct image-to-code & 62.80 & -- & -- \\
\bottomrule
\end{tabular}
\end{table}

The Qwen3-VL reference uses the same 2,000 ChartQA examples and checkpoint as chart evidence capture. It receives the original chart and question and returns a program, which the shared local parser executes and scores.

\begin{table}[!t]
\centering
\setlength{\tabcolsep}{4pt}
\caption{Accuracy and leakage (\%) of local models, \textsc{ReCast}, and the remote solver.}
\label{tab:main-results}

\resizebox{\linewidth}{!}{%
\begin{tabular}{lrrrrrr}
\toprule
\multirow{2}{*}{Method}
& \multicolumn{3}{c}{Accuracy $\uparrow$}
& \multicolumn{3}{c}{Leakage $\downarrow$} \\
\cmidrule(lr){2-4}
\cmidrule(lr){5-7}
& ChartQA & NMSQA & Overall
& ChartQA & NMSQA & Overall \\
\midrule

\multicolumn{7}{l}{\emph{Local, original input}} \\
Qwen2.5-Omni-3B
& 23.70 & 46.30 & 35.00 & 0 & 0 & 0 \\
Gemma-4-E4B
& 49.50 & 44.60 & 47.05 & 0 & 0 & 0 \\
Qwen3-VL-8B
& 62.80 & -- & -- & 0 & 0 & 0 \\

\midrule
\textbf{\textsc{ReCast} (Ours)}
& 70.20 & 80.00 & 75.10 & 8.00 & 7.90 & 7.95 \\

\midrule
\multicolumn{7}{l}{\emph{Remote solver (Qwen3.5-Omni-Plus)}} \\
Remote-native, original media
& 75.10 & 87.40 & 81.25 & -- & -- & -- \\

\bottomrule
\end{tabular}%
}
\end{table}

The local rows establish the zero-leakage reference because their original inputs never leave the device. Remote-native serves as the full-exposure reference by construction: the original media and query are sent directly to the solver, and its accuracy therefore defines the unprotected reference for NAcc. Under our exposure definition, Remote-native reveals the complete source input by design, although we do not separately score it with the leakage judge.

\subsection{Release Checks}
\label{app:release-checks}

The protected-record check verifies that rendered values, labels, units, category-series associations, and queried visual attributes agree with the protected record and validated ChartSpec. For speech, it verifies that the transcription of the candidate waveform agrees with the protected passage in quantities, entity references, and the content the query needs. The source check aligns rewritten entities with their source roles, inverts mapped values, and compares the result with the original input. For charts, it covers category-value assignments, legend-series associations, ordering, and queried visual attributes; for speech, it covers entity-quantity associations, coreference, event order, and temporal or comparative dependencies. A candidate is released only if both checks pass.

\subsection{Text Anonymizer Diagnostics}
\label{app:anonymizer-diagnostics}

We adapt RLAA~\citep{duan2026rlaa} and
Eternis-Anonymizer-4B~\citep{eternis2025anonymizer} to the same
five-field record interface.

The source $A$ is the normalized five-field record, and $B$ is the anonymizer's output before numerical mapping. The judge is Qwen3.7-Plus~\citep{qwenteam2026qwen37} with temperature 0 and a 512-token budget. Coverage is the fraction of records accepted by the adapter. Leakage is the fraction of positive judgments among accepted records with resolved judgments.

\begin{table}[H]
\centering\scriptsize
\setlength{\tabcolsep}{3pt}
\caption{Per-modality coverage and leakage.}
\label{tab:b-2}
\resizebox{\linewidth}{!}{%
\begin{tabular}{lrrrr}
\toprule
Rewriter & ChartQA coverage & ChartQA leakage & NMSQA coverage & NMSQA leakage \\
\midrule
RLAA-Qwen3-4B & 100.00 & 86.80 & 100.00 & 73.70 \\
Eternis-Anonymizer-4B & 100.00 & 96.70 & 100.00 & 93.20 \\
\bottomrule
\end{tabular}}
\end{table}

Both adapters use the five-field record. RLAA uses Qwen3-4B, up to three
revisions, and numerical-literal checks; Eternis uses its official
opening-bracket prefill, with quantity replacements removed to meet the
numerical-literal contract.

\subsection{Human Calibration of the Leakage Judge}
\label{app:human-calibration}

\textbf{Sample.} We drew 400 solver-bound pairs in eight cells of 50. Six in-distribution cells cross the two datasets (ChartQA, NMSQA) with three Rewriter conditions (untuned Qwen3-4B, distilled \textsc{ReCast}, and the GPT-5.6-Sol teacher). Two out-of-distribution cells contain distilled \textsc{ReCast} outputs on ChartX and Spoken-MQA, so that calibration also covers the setting of Tab.~\ref{tab:3}. Judge-positive and judge-negative cases were sampled in roughly equal numbers within each cell; cells with few judge-positive cases, such as the teacher on NMSQA, include all of them.

\textbf{Annotation.} Three annotators labeled each pair with the four-view rubric of Appendix~\ref{app:prompts}. Pairs were shuffled across cells. Annotators saw the original context A and the solver-bound context B, and they did not see the condition or the judge's labels. Disagreements were resolved by majority vote; for the any-positive decision, each annotator's label is the OR of their four views. Manual inspection of calibration cases containing preserved integers
in the 0-15 range did not identify judge errors attributable to this
mapping rule.

\begin{table}[H]
\centering\scriptsize
\setlength{\tabcolsep}{3pt}
\caption{Agreement between the judge and human labels.}
\label{tab:b-3}
\resizebox{\linewidth}{!}{%
\begin{tabular}{lrrr}
\toprule
View & Judge-human agreement (\%) & Cohen's $\kappa$ (judge vs. human) & Inter-annotator $\kappa$ \\
\midrule
Semantic & 92.0 & 0.82 & 0.81 \\
Lexical & 89.5 & 0.74 & 0.72 \\
Correspondence & 93.8 & 0.85 & 0.84 \\
Linkability & 88.3 & 0.70 & 0.69 \\
Any-positive (reported metric) & 91.5 & 0.83 & 0.78 \\
\bottomrule
\end{tabular}}
\end{table}

On the any-positive decision, the judge labels 202 pairs positive and the human majority labels 198 positive, with 183 in common. The judge's precision and recall against human labels are therefore 90.6\% and 92.4\%. Reweighting the two in-distribution \textsc{ReCast} cells to the full evaluation set gives a human-labeled leakage rate of 8.70\%, compared with 7.95\% from the judge.

\subsection{PSIC Diagnostic}
\label{app:psic}

PSIC~\citep{shen2025psic} produces a protected decoding of each of 2,000
ChartQA images. Our diagnostic compares the protected image with its source
using the native multimodal Qwen3.5-Omni-Plus~\citep{qwenteam2026qwen35omni}
leakage judge, without the
question or a downstream solver request. All 2,000 judgments resolve, with
564 positive decisions (28.20\%). This image-only measure does not include
the full solver-bound context used to score \textsc{ReCast}. PSIC does not provide the operand-recovery mechanism required to recover
the original numerical answer, so its protected outputs cannot support
end-to-end answering under our protocol.

\subsection{SPTI Diagnostic}
\label{app:spti}

SPTI~\citep{wang2025spti} synthesizes 2,000
ChartQA images using local Qwen3-4B and
$(\epsilon,\delta)=(1.0,1.448\times10^{-4})$ (clean-image FID 89.70).
Across the dataset, no source-content leakage is detected
under the caption-based diagnostic. Shuffled voting leaves the synthesized images unpaired with the original
questions, so they cannot be used for sample-level end-to-end answering
under our protocol.

\subsection{AudioShield Diagnostic}
\label{app:audioshield}

AudioShield~\citep{jin2025whispering} protects 2,000 NMSQA recordings with
the released LS-TUAP algorithm path and an upstream VITS backend at 16\,kHz.
The run manifest marks this configuration as non-paper-equivalent. We compare
the original normalized transcript with Qwen3-ASR-1.7B~\citep{shi2026qwen3asr}
transcriptions of the protected waveforms. The Qwen3.7-Plus~\citep{qwenteam2026qwen37} text leakage
judge resolves all 2,000 pairs and labels 12 positive (0.60\%). The
protected-audio transcriptions are at most 20 characters for 1202 examples;
the low leakage rate therefore also reflects suppression of recognized
speech. AudioShield is designed to suppress recognizable speech rather than
preserve the task content required for numerical reasoning; its protected
audio therefore cannot support end-to-end answering under our protocol.

\section{Hyperparameters}
\label{app:hyperparameters}

The student is Qwen3-4B~\citep{yang2025qwen3} trained with
QLoRA~\citep{dettmers2023qlora}; the offline teacher is
GPT-5.6-Sol~\citep{openai2026gpt56}.

\begin{table}[H]
\centering\small
\setlength{\tabcolsep}{3pt}
\caption{Rewriter training.}
\label{tab:c-1}
\begin{tabular}{lr}
\toprule
Parameter & Value \\
\midrule
Student & Qwen3-4B, QLoRA (NF4, bfloat16 compute) \\
LoRA rank / $\alpha$ / dropout & 16 / 32 / 0.05, all linear layers \\
Learning rate / schedule & 1e-4, cosine, 0.03 warm-up ratio \\
Epochs / gradient accumulation & 2 / 8 \\
Maximum sequence length & 4,096 \\
Loss & assistant tokens only \\
Optimizer & paged AdamW 8-bit \\
Checkpoint interval / seed & 250 steps / 42 \\
Final validation loss & 0.404 \\
Teacher & GPT-5.6-Sol, temperature 0, max 5,000 output tokens, up to 3 retries \\
\bottomrule
\end{tabular}
\end{table}

\begin{table}[H]
\centering
\caption{Inference settings.}
\label{tab:c-2}
\footnotesize
\renewcommand{\arraystretch}{1.0}
\setlength{\tabcolsep}{4pt}

\begin{tabularx}{\linewidth}{
    >{\raggedright\arraybackslash}p{0.28\linewidth}
    >{\raggedright\arraybackslash}X
}
\toprule
Component & Setting \\
\midrule
Chart-to-text & greedy decoding, bfloat16, max 2,200 new tokens \\
Speech-to-text & bfloat16, batch size 1, max 512 new tokens \\
Rewriter & thinking disabled, temperature 0.2, top-$p$ 0.9, max 16,384 new tokens \\
Numerical map & policy \texttt{preserve\_integers\_0\_15\_semantic\_years.v1}, seed 42 \\
ChartSpec generation & JSON output, max 6,000 tokens, up to 3 attempts \\
ChartSpec review & max 2,000 tokens, up to 3 attempts \\
Chart repairs / render timeout & 4 / 45 s \\
TTS & speed 1.0, 120 s timeout \\
Solver & temperature 0, max 2,000 tokens, streaming; up to 5 transport attempts and 3 format repairs \\
Leakage judge & temperature 0, max 512 tokens, up to 2 transport retries and 1 format repair \\
Program execution & separate process, 5 s timeout \\
\bottomrule
\end{tabularx}
\end{table}

\section{Prompt Specifications}
\label{app:prompts}
\begingroup
\small
\setlength{\parskip}{0pt}

\textsc{ReCast} uses task-specific prompts for evidence capture, semantic rewriting,
remote reasoning, and source-content auditing. The prompts are designed to
make the role of each model explicit while keeping the protection and
recovery logic independent of model-specific hidden representations.

\paragraph{Chart evidence capture.}
The chart-to-text model receives the chart together with its query and is
instructed to extract the evidence required for downstream reasoning rather
than answer the query directly. The prompt requests visible labels, values,
units, category-value and legend-series correspondences, ordering, and
query-relevant visual attributes. The resulting description is organized as
task evidence for the shared textual record.

\paragraph{Joint semantic rewriting.}
The Rewriter operates on the complete five-field record and rewrites the
evidence and query jointly into a coherent but semantically distinct domain.
It is instructed to change source-specific entities, topics, and descriptions
while preserving evidence-query correspondence and task-relevant relations.
These include quantity roles, category-value and legend-series associations,
coreference, event order, temporal or comparative relations, the requested
reasoning operation, and the answer type. Numerical literals must remain
unchanged at this stage and are handled separately by the numerical mapping
module.

\paragraph{Remote solver.}
The solver receives only the protected artifact and protected query. Its
prompt asks it to identify the evidence required by the query, select the
corresponding operands and operations, and return a single executable Python
program. Numerical operands read from the protected input are explicitly
marked as \texttt{protected(<number>)} so that they can be restored locally
before execution. The prompt also requires the solver to report when the
necessary evidence is unavailable rather than infer or fabricate missing
values.

\paragraph{Source-content audit.}
The audit prompt compares the original local context with the complete
solver-bound context and tests whether sample-specific source information
remains recoverable. It separates source-content leakage into semantic,
lexical, correspondence, and linkability views, while explicitly excluding
generic reasoning structure such as arithmetic, comparison, ordering, and
selection. The fixed audit prompt is reproduced below.

\begin{Verbatim}[
    fontsize=\scriptsize,
    breaklines=true,
    breakanywhere=true,
    breaksymbolleft={},
    breaksymbolright={},
    frame=single,
    framesep=1mm
]
Given original local multimodal context A and the complete solver-bound
context B, including retries and format repairs, audit B as a model-based
attack probe against A.

Return four labels: semantic_leakage, lexical_leakage, correspondence,
and linkability. For semantic_leakage, test leakage of original entities,
topics, categories, labels, values, dates, visual scenes, audio events, or
other sample-specific meaning. For lexical_leakage, test wording, phrase
fragments, local syntax, unusual n-grams, and distinctive expressions. For
correspondence, test whether original numerical values or entity-value
relations can be recovered or matched. For linkability, test whether B can
be linked or re-identified back to A from its protected content or
combinations of clues.
Ignore differences in modality, formatting, layout, sentence structure, and
ordinary surface wording. A generic operation such as counting, comparison,
ordering, arithmetic, ratio, selection, or extraction is not leakage by
itself when sample-specific content has changed. Compare B with A itself,
never with another rewritten sample.
Output exactly:
Semantic: Yes or No;
Lexical: Yes or No;
Correspondence: Yes or No;
Linkability: Yes or No;
Explanation: <brief evidence for each label>.
\end{Verbatim}
\endgroup

\clearpage
\section{Visualization Results}
\label{app:visualization-results}

\newcommand{\viscasepdf}[1]{\includegraphics[width=0.88\linewidth]{figure/visualization_cases/#1.pdf}}

\begin{figure}[H]
\centering
\viscasepdf{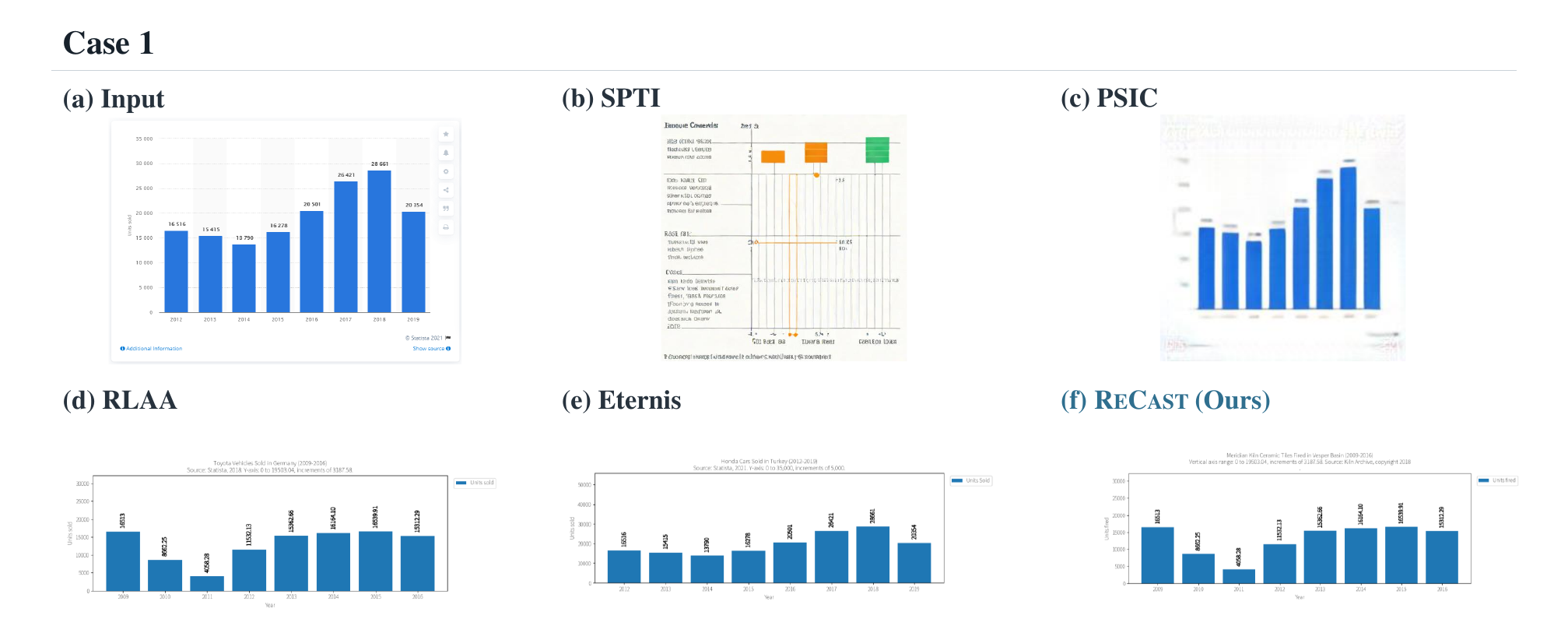}\par\vspace{2pt}
\viscasepdf{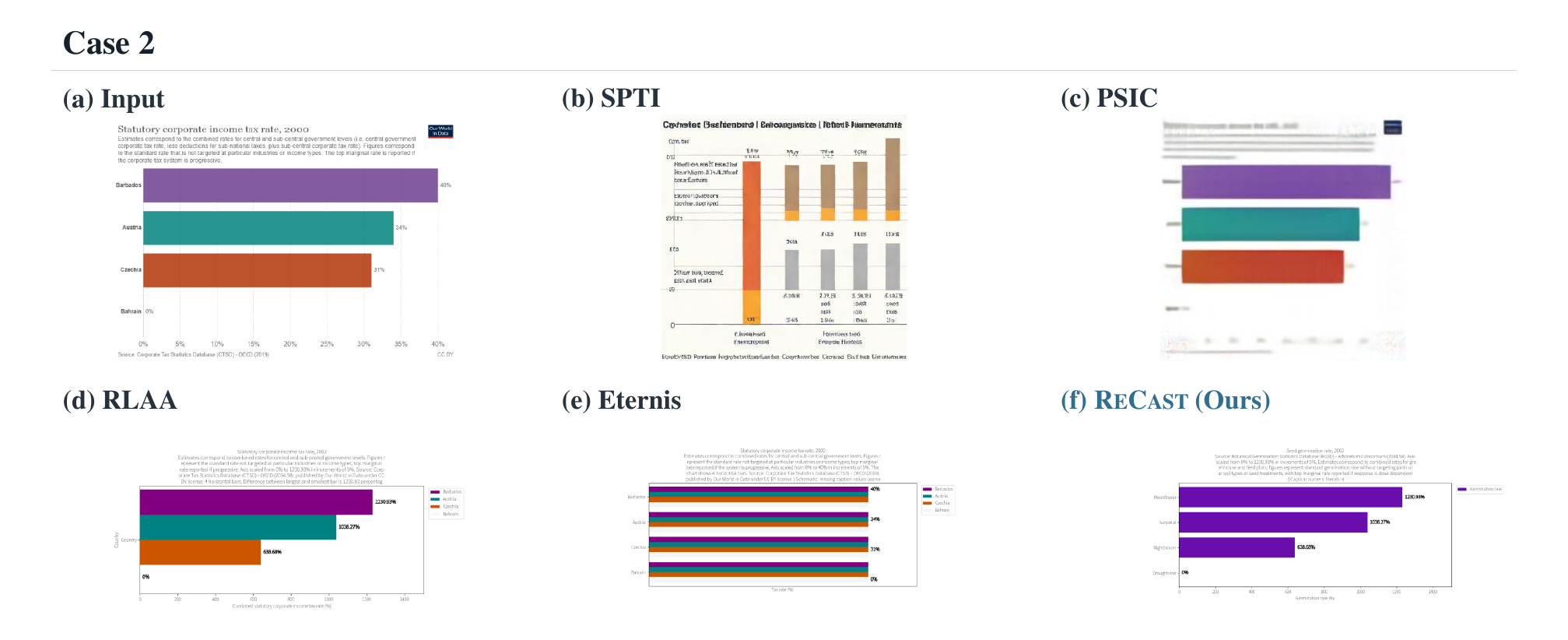}\par\vspace{2pt}
\viscasepdf{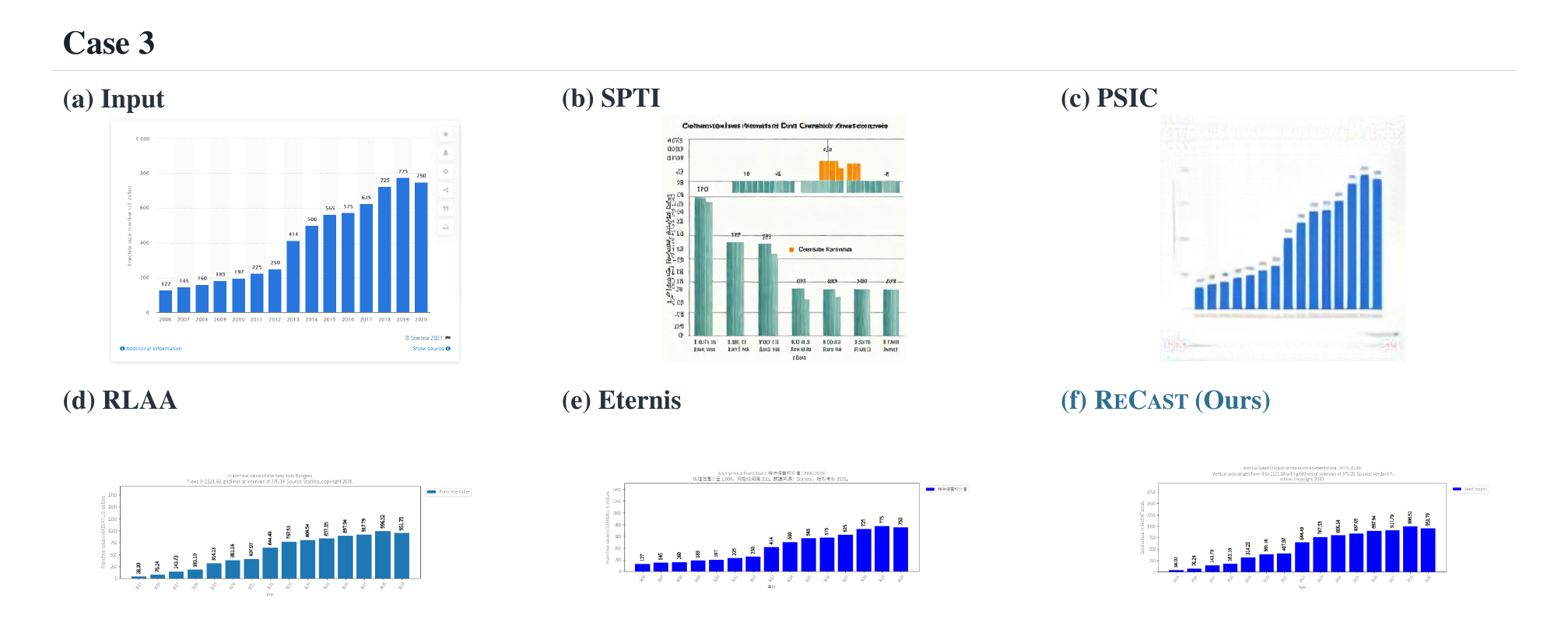}\par\vspace{2pt}
\viscasepdf{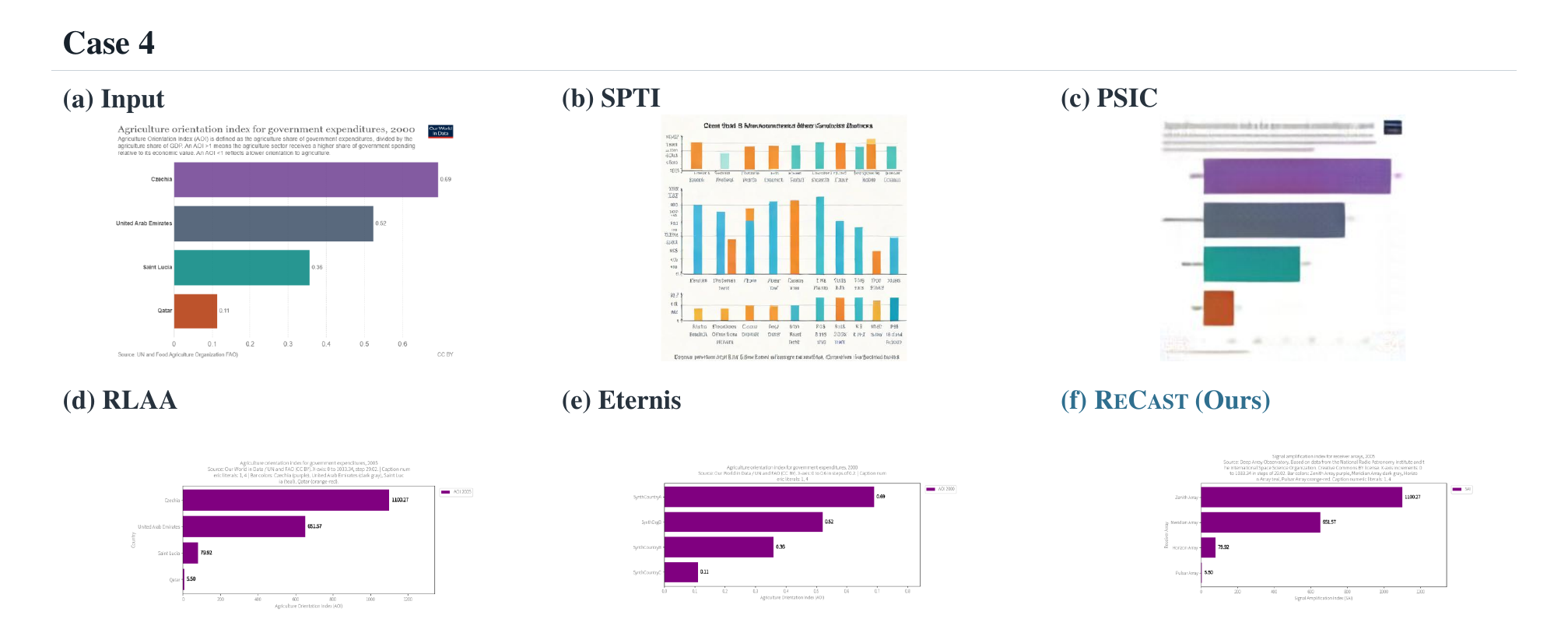}
\caption{Four ChartQA comparisons: input, SPTI~\citep{wang2025spti},
PSIC~\citep{shen2025psic}, RLAA~\citep{duan2026rlaa},
Eternis~\citep{eternis2025anonymizer}, and \textsc{ReCast}. SPTI
samples are unpaired; the other outputs follow the source sample ID.}
\label{fig:visualization-images}
\end{figure}

\clearpage
\begin{figure}[H]
\centering
\viscasepdf{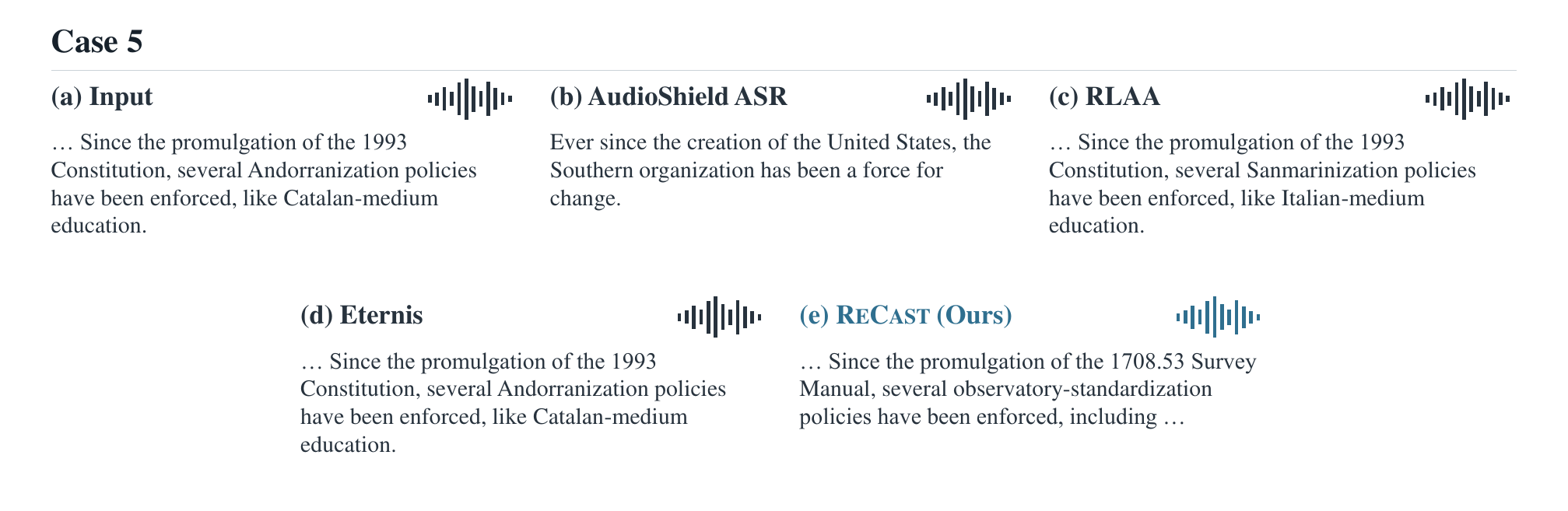}\par\vspace{2pt}
\viscasepdf{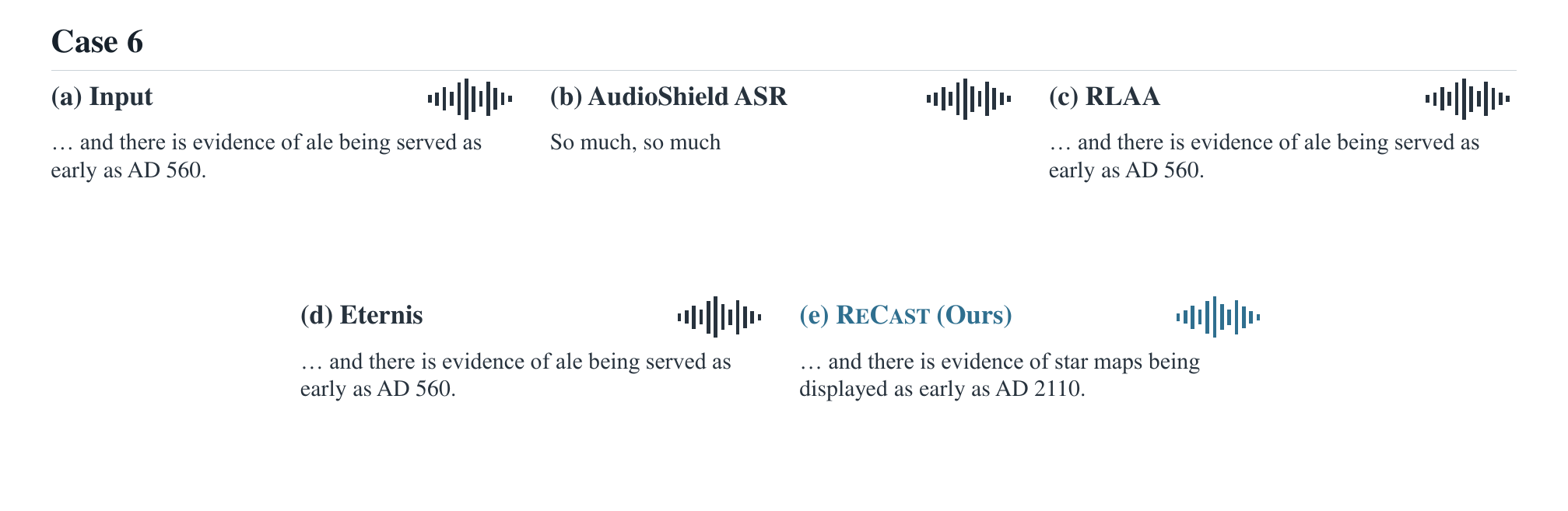}\par\vspace{2pt}
\viscasepdf{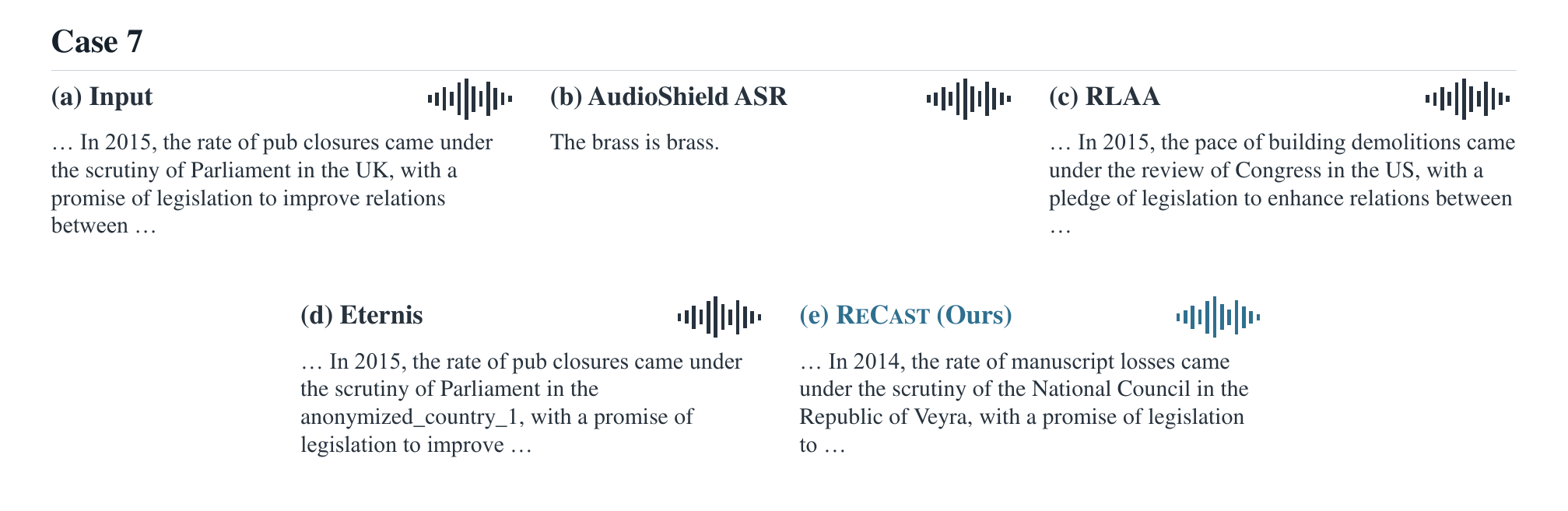}\par\vspace{2pt}
\viscasepdf{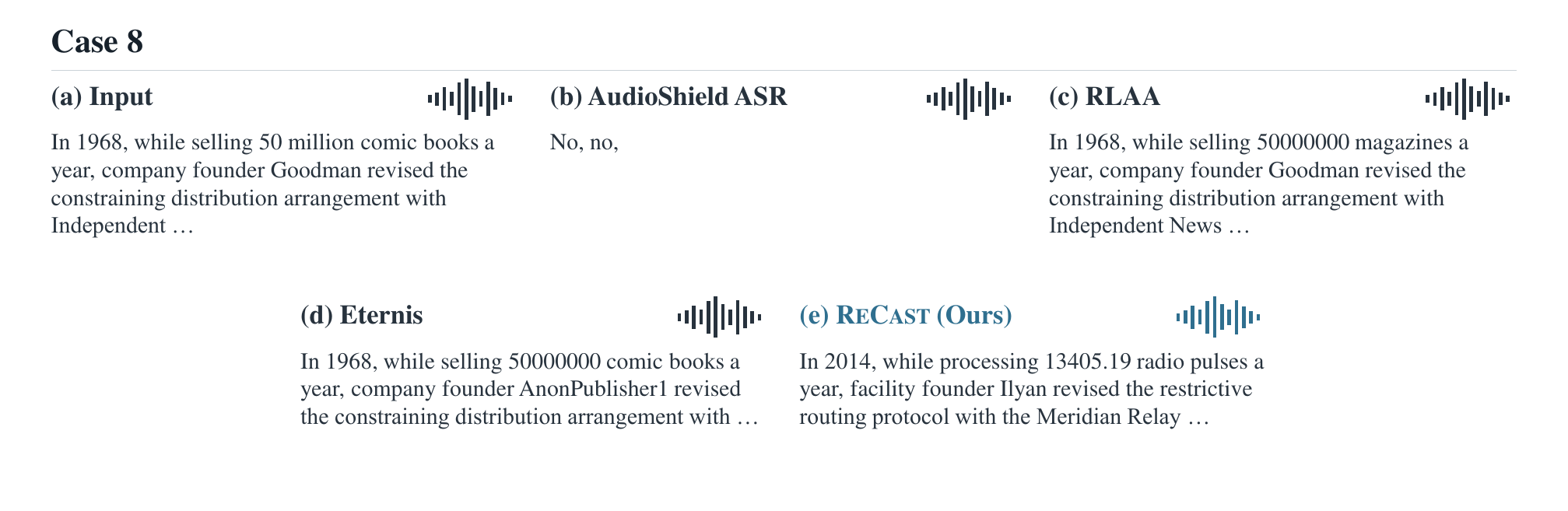}\par\vspace{2pt}
\viscasepdf{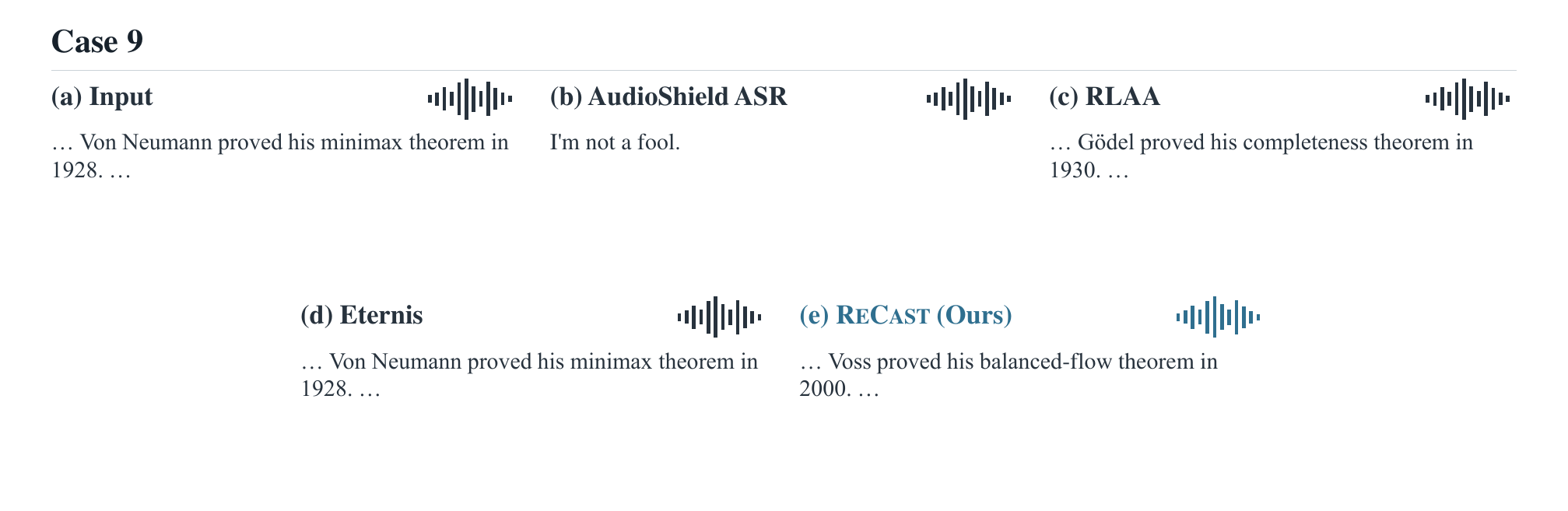}
\caption{Five NMSQA audio comparisons: input transcript,
AudioShield~\citep{jin2025whispering} protected-audio ASR,
RLAA~\citep{duan2026rlaa} and Eternis~\citep{eternis2025anonymizer}
protected text, and \textsc{ReCast} transcript.}
\label{fig:visualization-audio}
\end{figure}

\endgroup

\end{document}